\documentclass[Journal]{IEEEtran}
\IEEEoverridecommandlockouts

\usepackage{amssymb}
\usepackage[cmex10]{amsmath}
\usepackage{stfloats}
\usepackage{graphicx}
\usepackage{subfigure}
\usepackage{tabularx}
\usepackage{epsfig,epsf,color,balance,cite}
\usepackage{verbatim}
\usepackage{url}
\usepackage{bm}
\usepackage{booktabs}
\usepackage{float}
\usepackage{subfigure}
\usepackage{subcaption}
\usepackage{caption}
\usepackage{amsmath}
\allowdisplaybreaks[2]

\usepackage[ruled,linesnumbered]{algorithm2e}

\usepackage{algorithmic}

\usepackage{color}
\definecolor{myc1}{rgb}{0,0,0}

\begin{document}

\title{ 

Green Probabilistic Semantic Communication over Wireless Networks
}

\author{
\IEEEauthorblockN{Ruopeng Xu,
                  Zhaohui Yang,
                  Yijie Mao,
                  Chongwen Huang,
                  Qianqian Yang, 
                  Lexi Xu,
                  Wei Xu, ~\IEEEmembership{Senior Member,~IEEE,}
                  and Zhaoyang Zhang~\IEEEmembership{Senior Member,~IEEE,}
                 } 
 \thanks{R. Xu, Z. Yang, C. Huang, Q. Yang and Z. Zhangare with the College of Information Science and Electronic Engineering, Zhejiang University, and also with Zhejiang Provincial Key Laboratory of Info. Proc., Commun. \& Netw. (IPCAN), Hangzhou, 310027, China (e-mails: \{ruopengxu, yang\_zhaohui, hongwenhuang, qianqianyang20, ning\_ming\}@zju.edu.cn).}
\thanks{Y. Mao is with the School of Information Science and Technology, ShanghaiTech University, Shanghai 201210, China (e-mail:
maoyj@shanghaitech.edu.cn).}
\thanks{L. Xu is with the Research Institute, China United Network Communications Corporation, Beijing 100048, China (e-mail: davidlexi@hotmail.com).}

\thanks{W. Xu is with National Mobile Communications Research Laboratory, Southeast University, Nanjing, 211189, China (e-mail: wxu@seu.edu.cn).}

\vspace{-2em}
}

\maketitle

\begin{abstract}
In this paper, we propose a multi-user green semantic communication system facilitated by a probabilistic knowledge graph (PKG). By integrating probability into the knowledge graph, we enable probabilistic semantic communication (PSC) and represent semantic information accordingly. 
On this basis, a semantic compression model designed for multi-user downlink task-oriented communication is introduced, utilizing the semantic compression ratio (SCR) as a parameter to connect the computation and communication processes of information transmission. Based on the rate-splitting multiple access (RSMA) technology, we derive mathematical expressions for system transmission energy consumption and related formulations. Subsequently, the multi-user green semantic communication system is modeled and the optimal problem with the goal of minimizing system energy consumption comprehensively considering the computation and communication process under given constrains is formulated. In order to address the optimal problem, we propose an alternating optimization algorithm that tackles sub-problems of power allocation and beamforming design, semantic compression ratio, and computation capacity allocation. Simulation results validate the effectiveness of our approach, demonstrating the superiority of our system over methods using Space Division Multiple Access (SDMA) and non-orthogonal multiple access (NOMA) instead of RSMA, and highlighting the benefits of our PSC compression model.
\end{abstract}

\IEEEpeerreviewmaketitle

\section{Introduction} \label{Introduction}

Applications of next generation wireless communication technology, such as digital twin, extended reality, and self-driving car technology, require a large amount of data to be transmitted within a low time latency, which puts higher and higher requirements on the data transmission rate. Semantic communication is considered to be a new paradigm with great potential in the sixth generation wireless communication (6G) standard \cite{zhang2023model} to meet the requirements above. Different from the traditional communication paradigm proposed by Shannon in \cite{shannon1948mathematical} that aims to maximize the channel transmission rate in Shannon's information theory, semantic communication is task-oriented and focuses on the semantic information required by the specific communication task. Therefore, it can greatly reduce the amount of the data to be transmitted, which makes the research on semantic communication very valuable. Besides, the continuous development of artificial intelligence (AI) technology provides a strong boost to semantic communication. For instance, the development of natural language processing (NLP) technology enables computers to better understand and process human language, and promotes the continuous improvement of accuracy and completeness of semantic extraction methods such as text understanding, speech recognition and video processing. On the other hand, the development of semantic communication technology also promotes the advancement of AI systems in applications such as intelligent assistants, chatGPT and automatic translation. At the same time, since the wireless communication network environment is constrained by time delay, frequency, space, energy consumption, user service quality and other issues, comprehensive consideration of resource allocation in the multi-user communication transmission process under a wireless communication resource-constrained environment, establishing corresponding constrained optimization problems and obtaining the optimal solution are also topics that are greatly significant in the research of the new generation of communication systems. 

In 1948, Shannon published \cite{shannon1948mathematical}, marking the birth of information theory. The following year, in \cite{shannon1949mathematical}, Shannon explained the basic problems of communication, gave a general communication system model and defined the Shannon formula. Shannon and Weaver proposed three levels of communication in \cite{weaver1953recent}: the technical level, how to ensure the correct transmission of communication symbols; the semantic level, how to convey the exact meaning of the sent symbols; and the pragmatic level, how the received meaning affects the system's behaviors in the expected way. In the following decades, the development of communication technology has always revolved around the basic problem of communication: "how to increase the transmission rate as much as possible under the constraints of the channel", focusing on the error-free transmission of symbols at the bit level. However, with the development and iteration of wireless communication technology and the vigorous development of communication-related fields, data traffic has grown exponentially, which caused a heavy burden and brought a huge challenge on the conventional communication.

In recent years, semantic communication technologies based on wireless networks have developed in a abundent variety\cite{xu2023edge}. \cite{uysal2022semantic} uses the importance of semantic information to construct and evaluate communication networks. \cite{kountouris2021semantics} comprehensively considers process dynamics, signal sparsity, data correlation and semantic attributes and uses goal-oriented semantic communication for information generation, transmission and reconstruction, and applies it to real-time source reconstruction to achieve remote driving scenarios. According to the different semantic source modalities in different tasks, different deep learning models for semantic communication have been studied in it. \cite{xie2021deep} proposed a semantic communication model (DeepSC) based on deep learning to optimize the mutual information between the original information and the encoded signal, which can jointly perform semantic communication channel coding for text transmission. Based on the DeepSC model, \cite{xie2020lite} designed a distributed semantic communication system (L-DeepSC) for networks with limited device power and computing power, and used it for low-complexity text transmission. 

The use of deep learning technology to complete the extraction of semantic information is an important aspect of deep learning for semantic communication. Common semantic information extraction technologies include: word-of-bag-model\cite{zhang2010understanding}, representing the collection of texts as a collection of words and ignoring its order and grammar; word embedding model\cite{wang2019evaluating}, mapping words into a continuous vector space to capture the semantic and grammatical relationships between words.; topic model\cite{vayansky2020review} , used to discover the hidden topic structure in text, thereby performing semantic extraction; and deep learning models such as, recurrent neural network (RNN), long short-term memory network (LSTM), and Transformer model, used to learn abundant semantic representations in text.

The development of deep learning technology has driven the continuous improvement and advancement of knowledge graphs (KGs) and related technologies. KGs are structured representations of abstract facts, which contain a variety of entities and relationships between entities. They have high information density and are one of the important ways to represent knowledge in the big data era. In essence, they can be understood as a large-scale semantic network. This property makes KGs and semantic communication have a natural connection. Their development and the continuous maturity of technology have brought possibilities for semantic extraction and representation of semantic communication. \cite{bordes2014semantic} proposed an energy method for semantic matching, which maps the relationship between entities to the vector of the input layer and relies on neural networks for semantic matching. \cite{wang2017knowledge} proposed embedding entities and relationships in KGs into dense low-dimensional vector spaces, simplifying downstream operations while maintaining the inherent structure of KGs. \cite{jiang2022reliable} uses KGs to convert sentences to be transmitted into triples, sorts them according to semantic importance, and adaptively transmits content according to channel quality, allocating more transmission resources to important triples to enhance communication reliability. \cite{wang2023knowledge} proposed to use facts in the knowledge base for semantic reasoning and decoding, and used the public WebNLG corpus for simulation, which demonstrates a good performance.

However, previous studies have shown low cross modal applicability of communication framework modes, insufficient consideration of resource constraints in wireless communication, and insufficient consideration of the increased energy consumption after introducing computation. For example, both \cite{uysal2022semantic} and \cite{kountouris2021semantics} have constructed their respective semantic communication frameworks for transmitting control signals in real-time control systems as semantic messages. However, the semantic information in these frameworks cannot represent the semantics of messages such as text, images, or videos; The deep learning models in \cite{xie2020lite,xie2021deep,dai2022nonlinear,dong2022semantic,wang2022wireless,jiang2022wireless} only work for specific sources, and a change in the modality of the source means that the model is no longer applicable. \cite{bao2011towards,maatouk2022age,weng2021semantic,guler2018semantic} assume that all semantic information extracted from the original data can be transmitted through the network, ignoring the practical problem of resource constraints in infinite communication such as bandwidth, power, and latency; At the same time, research on resource allocation in some existing wireless networks is not applicable for semantic communication. \cite{bordes2014semantic,wang2017knowledge,jiang2022reliable,wang2023knowledge} did not fully consider the computational energy consumption brought about in the creation, use, and updating of KGs. In the face of massive data and huge energy consumption in communication systems today, it is not possible to fully utilize the energy input into the system. Meanwhile, it is unreasonable to directly apply KGs in the field of communication without any changes.

Semantic communication focuses on the accurate and efficient accomplishment of communication tasks at the semantic level rather than precise transmission at the bit level, enabling it to better adapt to communication tasks with increasing data volume and lower latency requirements. At the level of semantic information extraction and representation, KG is a commonly used approach. Based on this, how to consider the entire process of information transmission more comprehensively and apply more adaptive multi access technology to make the constructed semantic communication system more `green' is the underlying starting point of this design. In view of this, this paper conducts research on the construction and optimization of multi-user semantic communication systems using probability KGs. The key contributions of the paper are listed as follows: 
\begin{itemize}
\item We propose a probabilistic knowledge graph (PKG) aided semantic communication system model by designing a semantic compression process consisting of the compression, transmission and decompression process the semantic information extracted and represented by the PKG under a multi-user scenario. The parameter semantic compression ratio (SCR) is as well introduced as a variable that can be mapped to both computational overhead and communication overhead to jointly consider the energy consumption of computation and communication process. Simultaneously given the overlapping nature of tasks needed to be transmitted for multi-user, we introduce rate-splitting multiple access (RSMA) to further alleviate the total energy consumption of the proposed system to make it more green.
\item Based on the proposed multi-user green semantic communication system model, we formulate an optimization problem to minimize system energy consumption by comprehensively considering computation and communication. On the basis of solving the power allocation and beamforming design sub-problem, the senmantic compression ratio sub-problem and the computation capacity allocation sub-problem respectively, we propose an alternating optimization algorithm to obtain an equivalent optimal solution to the original problem by iteratively and alternately optimizing the three sub-problems. 
\item Simulations on the system energy consumption of the three sub-problems, respectively, before and after the optimization process are launched firstly, verifying the correctness of the solutions to the sub-problems, which can further ensure the correctness of the solution to the original optimal problem. Subsequently, in the case of optimal resource allocation, we compared the total energy consumption of the system under different communication environments with or without the introduction of the semantic compression model to illustrate the effectiveness of the introduction of the semantic compression process in reducing system energy consumption. Last but not least, simulations on the total energy consumption of the system under different the multiple access modes, such as space division multiple access (SDMA) and non-orthogonal multiple access (NOMA) are carried out under different communication conditions to illustrate the adaptability of the multiple access mode to the proposed semantic communication system.
\end{itemize}

The rest of this paper is organized as follows. The specific design of the green probabilistic semantic communication system model is illustrated in Section \ref{Sec2}. Analysis and algorithm designs of addressing the formulated optimal problem and its according three sub-problems are presented in Section \ref{Algorithm Design}. Simulation process and outcomes are in the Section \ref{Simulation Results and Analysis}. Conclusions are drawn in Section \ref{Conclusion}.

\section{System Model Design} \label{Sec2}
\begin{figure*}[t]
\centering
\includegraphics[width=1\linewidth]{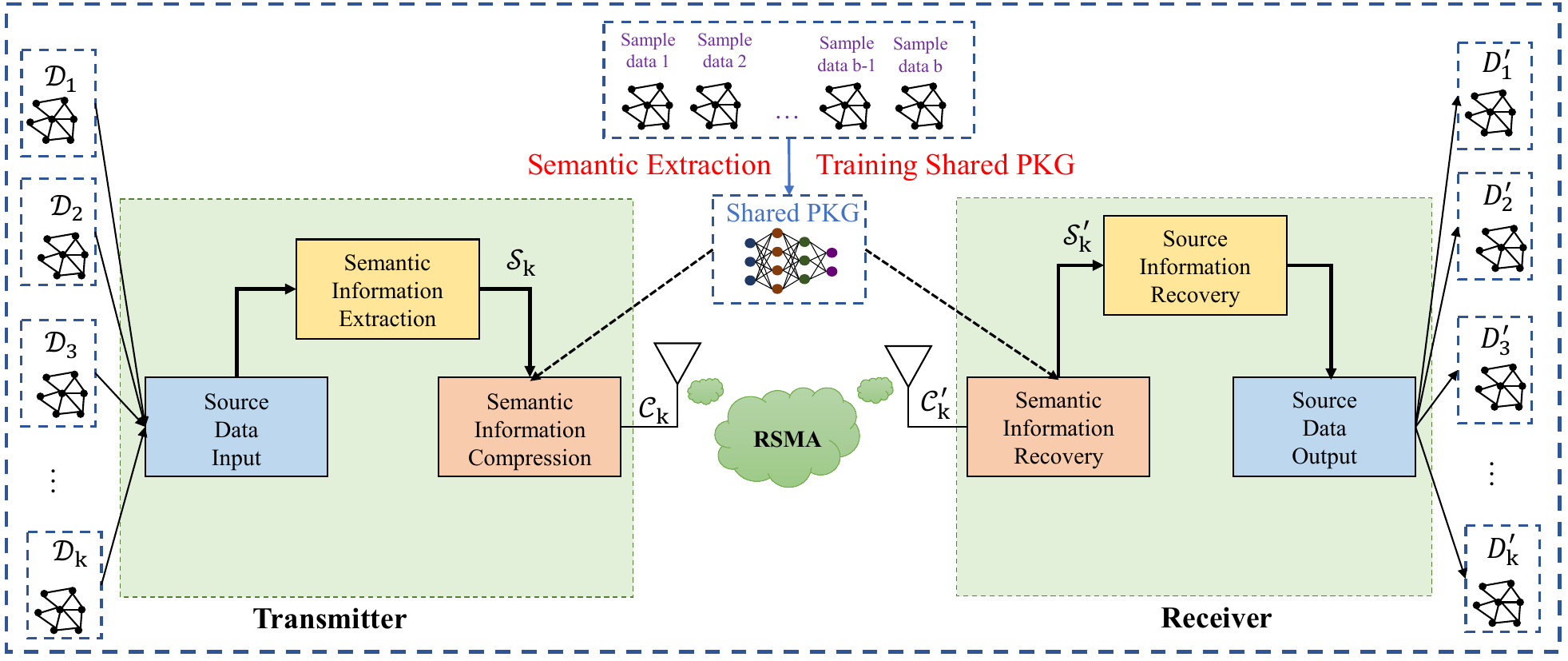}
\caption{System model design of PSC aided multi-user green semantic communication} 
\label{System Model Design}
\end{figure*}

With the gradual development of machine learning technology and large language models, great progress has been made in using triples to represent semantic information. However, from the perspective of the communication field, how to use triples representing semantic information to convey the messages with lower power consumption is an interesting and meaningful topic. In this section, a green probabilistic multi-user semantic communication system, as shown in Fig.~\ref{System Model Design}, is proposed with the goal of alleviating the total system consumption as much as possible. In Fig.~\ref{System Model Design}, the source data are first extracted to be semantic information, and then undergo the process of compression, transmission with RSMA and recovery with the help of the shared PKG and and then recovered to output containing the same semantic information at the receiver . In the rest of this section, we first clarify the definition of PKG, and the tell the process on how to extract semantic information, compress them at transmitter, transmit them and recover them at the receiver side. 

\subsection{Probabilistic Knowledge Graph}
Before introducing system model design, it is necessary to firstly clarify the concept of PKG.

\begin{figure}[t]
\centering
\includegraphics[width=1\linewidth]{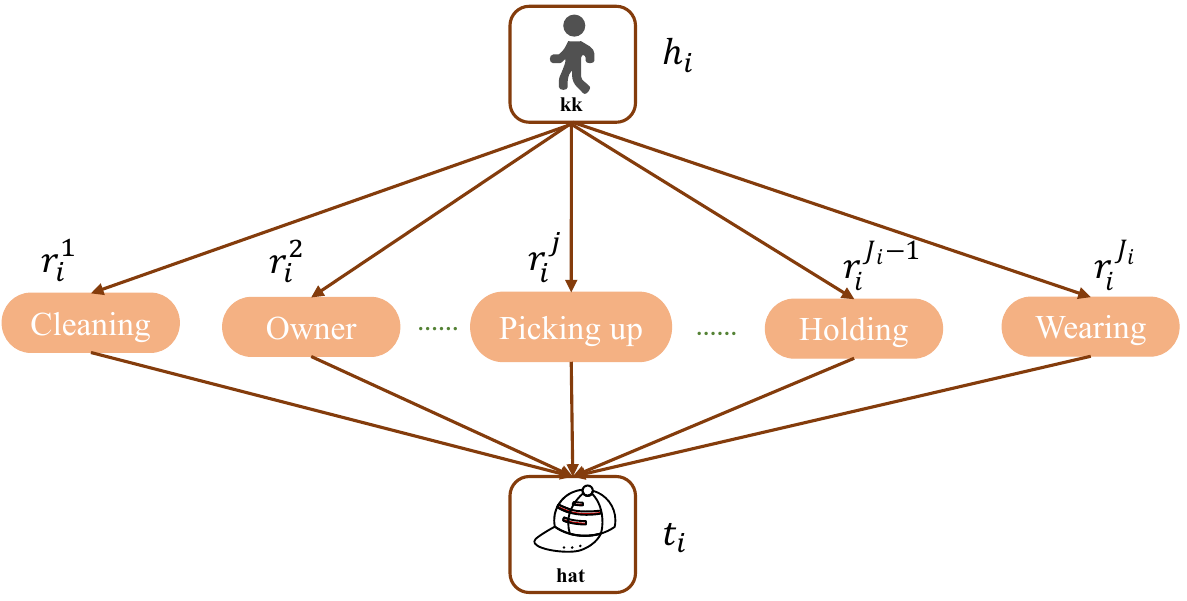}
\caption{A sample of an entity pair containing multiple relations} 
\label{notSingleRelation}
\end{figure}

It is assumed that the local KG has been extracted in advance at the base station (BS) with the sample data set $\mathcal{SD}$:
\begin{equation}
    \begin{aligned} 
    \mathcal{SD} = \{SD_1, SD_2, \dots, SD_Z\},
    \end{aligned}
\label{sample data set}
\end{equation}
where $SD_i$ stands for the sample data $i$ used to extract KG by the BS and $Z$ means there are $Z$ samples in the set utilized to extract the local KG. The basic elements of the local KG, which are also the semantic information of the semantic communication, are the triples\cite{xu2024resource} in the form of:
\begin{equation}
    \begin{aligned} 
    \varepsilon_i^j = (h_i,r_i^j,t_i),
    \end{aligned}
\label{Triple}
\end{equation}
where $\epsilon_i^j$ stands for the triple consisting of the entity pair $h_i$ and $t_i$ with the relation $r_i^j$ pointing from $h_i$ to $t_i$. The subscript $i$ is used to distinguish among different entity pairs, while the superscript $j$ is used to distinguish among different relationships within the same entity pair. What we need to clarify is that, as Fig.~\ref{notSingleRelation} showing, for one specific entity pair $h_i$ and $t_i$, there are maybe $J_i$ relations, from $r_i^1$ to $r_i^{J_i}$, i.e., more than one relation.

\begin{figure}[t]
\centering
\includegraphics[width=1\linewidth]{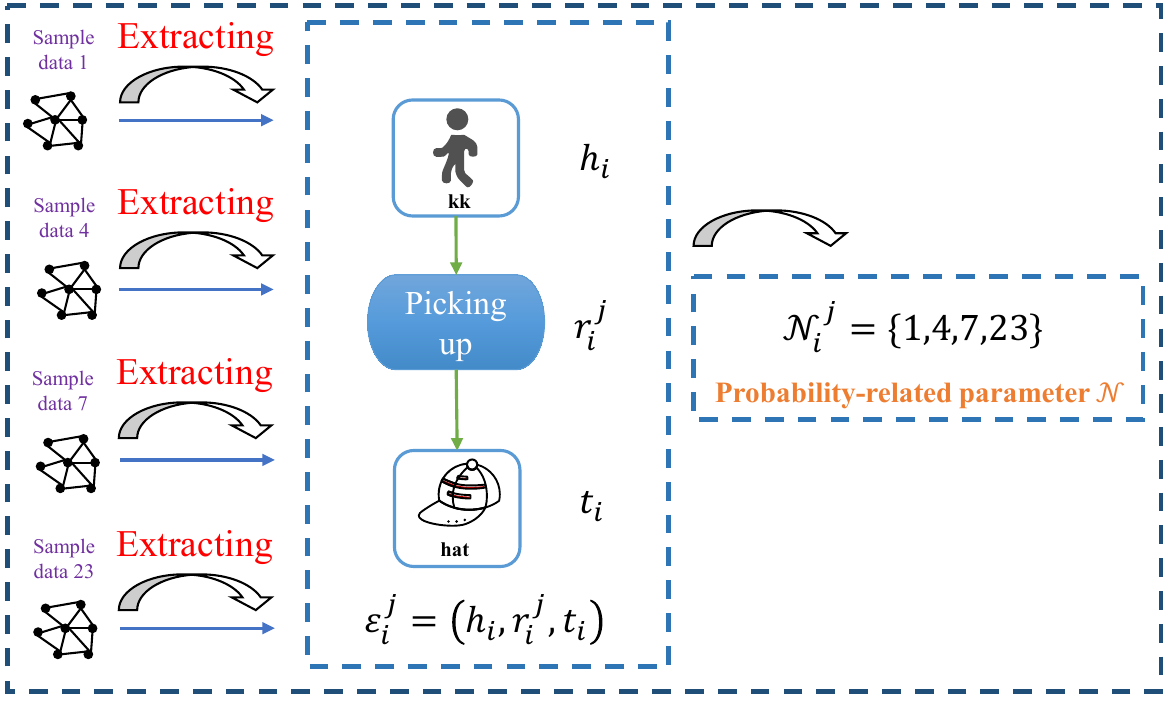}
\caption{Schematic diagram of probability-related parameter $\mathcal{N}$} 
\label{parameterN}
\end{figure}

Based on the definition of triples, if introduce the probability-related parameter $\mathcal{N}$ into the triples, one quadruple can be established in the form of:
\begin{equation}
    \begin{aligned} 
    \delta_i^j = (h_i,r_i^j,t_i,\mathcal{N}_i^j),
    \end{aligned}
\label{Quadruple}
\end{equation}
where $\mathcal{N}_i^j$ is a number set consisting of the serial number of data samples that can extract the triple $\varepsilon_i^j$. For example, as shown in Fig.~\ref{parameterN}, if semantic information $\varepsilon_i^j$ can be extracted from sample data ${SD}_1$, ${SD}_4$, ${SD}_7$ and ${SD}_{23}$, the number set $\mathcal{N}_i^j = \{1,4,7,23\}$.

On the basis of the definition of $\mathcal{N}$, the probability of a triple $\varepsilon_i^j$ in the KG can be defined as:
\begin{equation}
    \begin{aligned} 
    \kappa(\varepsilon_i^j) = \frac{\text{card}\left(\mathcal{N}_i^j\right)}{\sum_{l=1}^{J_i} \text{card}\left(\mathcal{N}_i^l\right)} = \kappa_i^j,
    \end{aligned}
\label{ProbabilityDefination}
\end{equation}
where $\text{card}(\mathcal{A})$ stands for the operation of calculating the number of the elements of set $\mathcal{A}$, $J_i$ demonstrates that there are $J_i$ relations in total of the given entity pair $h_i$ and $t_i$.

Further, if $\left(\cup_{a=1}^A \mathcal{N}_b^a\right) \cap \mathcal{N}_i^j \neq 0$ the conditional probability of $\varepsilon_a^b$ when known $\varepsilon_i^j$ can be defined as \eqref{ConditionalProbabilityDefination} shows.
\begin{equation}
    \begin{aligned} 
    \kappa(\varepsilon_b^a|\varepsilon_i^j) = \frac{\text{card}\left(\mathcal{N}_b^a \cap \mathcal{N}_i^j \right)}{\text{card}\left[\left(\cup_{a=1}^A \mathcal{N}_b^a\right) \cap \mathcal{N}_i^j\right]} = \kappa_{i,a,b}^j,
    \end{aligned}
\label{ConditionalProbabilityDefination}
\end{equation}
where $A$ is the total number of the relations pointing from $h_b$ to $t_b$ in the local KG. Otherwise, $\kappa_{i,a,b}^j = 0$. 

By expanding the triples of the KG to quadruples, the concept of constructing PKG on the basis of conventional knowledge graph is completed and shared PKG between BS and users can be generated accordingly.

\subsection{Semantic Information Extraction}
Consider the situation in which the BS has already generated a shared PKG before. The downlink is composed of a single BS with $L$ antennas and $K$ users each with a single antenna, and each user $k$ has huge data $\mathcal{D}_k$ waiting to be sent by the BS and to receive. The semantic information of each data set $\mathcal{D}_k$ are extracted based on the locally obtained knowledge base.

Source data to be transmitted can be represented as 
\begin{equation}\label{source data}
    \mathcal D_k =
    \{D_{k1}, D_{k2}, \cdots, D_{kU_k}\},
\end{equation}
where $\mathcal{D}_k$ is the source data of user $k$, $D_{kl}$ represents the $l$-th piece of the source data $\mathcal{D}_k$ and $U_k$ stands for the fact that there are $U_k$ pieces of source data in total. After being extracted by the BS, the data to be received of user $k$, can be represented as $\mathcal{S}_k$,
\begin{equation}
    \begin{aligned} 
     & \mathcal{S}_k = \{\varepsilon_x^y|\ \varepsilon_x^y\text{\  can\ be\  extracted\  from\ }\mathcal{D}_k \}.
    \end{aligned}
\label{MultiuserSemanticInformationSet}
\end{equation}

Given the fact that there are triples all users need to receive, we first compare all triples of the $K$ users, find the triples shared by every user, and merge them into shared triples, denoted as $\mathcal{U}_0$,
\begin{equation}
    \begin{aligned} 
        & \mathcal{U}_0 = 
        \cap_{k=1}^K \mathcal{S}_k
    \end{aligned}
\label{SharedTripleSet}
\end{equation}

The rest triples stay still and are denoted as private triples. For user $k$, private triple set is $\mathcal{U}_k$,
\begin{equation}
    \begin{aligned} 
        & \mathcal{U}_k = 
        \mathcal{S}_k - \mathcal{U}_0
    \end{aligned}
\label{PrivateTripleSet}
\end{equation}

Shared triples and private triples form the semantic information set $\mathcal{S}$ need to transmit,
\begin{equation}
    \begin{aligned} 
        & \mathcal{S} = 
        \cup_{k=0}^K \mathcal{U}_k
    \end{aligned}
\label{SemanticTripleSet}
\end{equation}

\subsection{Semantic Information Compression}
Denote the set containing the triples sharing the same entity pair as one quadruple set,
\begin{equation}
    \begin{aligned} 
    \Delta_i = \bigg\{h_i,\Big[\big(r_i^1, \mathcal{N}_i^1 \big),\dots,\big(r_i^{J_i},\mathcal{N}_i^{J_i}\big)\Big],t_i\bigg\},
    \end{aligned}
\label{QuadrupleSet}
\end{equation}
where $\Delta_i$ is the quadruple set with entity pair $h_i$ and $t_i$, and $J_i$ relations between $h_i$ and $t_i$. 

Assume that there are $\Phi$ quadruple sets can be constructed from the local KG. Utilizing the definitions of the probability, we first calculate the probability of all the triples and form the probability matrix with no prior knowledge,
\begin{equation}
    \begin{aligned}
      \bm{\kappa} =
      \begin{bmatrix}
      \kappa_1^1 & \kappa_1^2 & \dots & \kappa_1^{J_1} & \dots & 0\\
      \vdots & & & & & \vdots\\
      \kappa_m^1 & \kappa_m^2 & \dots & \dots & \dots & \kappa_m^{J_m}\\
      \vdots & & & & & \vdots\\
      \kappa_\Phi^1 & \kappa_\Phi^2 & \dots & \kappa_\Phi^{J_\Phi} & \dots & 0\\
      \end{bmatrix}
      \in \mathcal{R}^{\Phi \times J_m},
    \end{aligned}
\label{ProbabilityMatrixwithnoKnowledge}
\end{equation}
where $\kappa_i^j$ means the probability corresponding to quadruple $\delta_i^j$ in $\Delta_i$, and $J_m = {\max\{J_i\}}_{i=1}^{\Phi}$. In order to ensure that each row of the probability matrix can represent a quadruple set, the rows corresponding to the quadruples with fewer relations are padded with zeros.

To reduce the communication overhead, we can omit the relation of the triple that is most likely to appear in a given entity pair, and a marker, $\phi_n$, with very little information can be used to indicate the round $n$ in which the relation of the triple is omitted. Denote a triple whose relation is replaced by the marker, as $o_i^n$,
\begin{equation}
    \begin{aligned} 
     & o_i^n = (h_i,\phi_n,t_i),
    \end{aligned}
\label{MultiuserSemanticInformation}
\end{equation} 
where $o_i^n$ is the triple degraded at round $n$.

For the semantic information set $\mathcal{S}$ in a specific transmission task, to find the triples which can be degraded, traverse every triple of the set. If the probability corresponding to an arbitrary triple $\varepsilon_i^j$ is the value of the matrix $\bm{\kappa}$ and this value is the maximum value of one row, $\varepsilon_i^j$ is supposed to be degraded to $o_i^n$. After the first round of traversal, some triples in the set $\mathcal{S}$ have completed degradation. Taking the triples degraded during the first traversal as condition, a one-dimensional conditional probability matrix can be calculated. For example, if there are three triples $\{\varepsilon_2^5, \varepsilon_7^3\}$ are judged to be degenerated in the first compression round, and they will be denoted as $\{o_2^1, o_7^1\}$. Taking $o_2^1$ as the condition, calculate the conditional probability of each triple in each quadruple set to obtain $\bm{\kappa}_1^{o_2^1}$, which can be expressed as:
\begin{equation}
    \begin{aligned}
      \bm{\kappa}_1^{o_2^1} =
      \begin{bmatrix}
      \kappa_{1,1,2}^1  & \dots & \kappa_{1,1,2}^{J_1} & \dots & 0\\
      \vdots & & & & \vdots\\
      \kappa_{m,1,2}^1  & \dots & \dots & \dots & \kappa_{m,1,2}^{J_m}\\
      \vdots & & & & \vdots\\
      \kappa_{\Phi,1,2}^1 & \dots & \kappa_{\Phi,1,2}^{J_\Phi} & \dots & 0\\
      \end{bmatrix}
    \end{aligned}
\label{OneDimentionalProbabilityMatrixwithnoKnowledge}
\end{equation}
where subscript $1$ and superscript $o_2^1$ of $\bm{\kappa}_1^{o_2^1}$ mean that it is a one-dimensional matrix and it takes $o_2^1$ as the prior knowledge, respectively. Then traverse the triples that are not degenerate. If the conditional probability value corresponding to a triple is in the matrix $\bm{\kappa}_1^{o_2^1}$ and is the maximum value of a row of the matrix, replace the relation by the token $\phi_2$ and record it as degenerate as well. Then generate new one-dimensional conditional probability matrices $\bm{\kappa}_1^{o_7^1}$ based on ${o_7^1}$, traverse the non-degraded triples with the same replacement principle, and the second compression round is completed.

Subsequently, by constructing two-dimensional probability matrices, three-dimensional probability matrices and even $Q$-dimensional probability matrices, the third, the fourth, and even the $(Q+1)$-th round of semantic information compression can be completed. It is obvious that if there are $n_1$ triples replaced in the first round, there will be $C_{n_1}^1$ one-dimensional matrices to be calculated in the second round, and without loss of generality, if there are $n_{Q}$ triples replaced in the $Q$-th round, there will be $C_{\sum_{i=1}^Q n_i}^{Q} $ matrices in $Q$-dimension to be calculated in the $(Q+1)$-th round.

While the operation to omit the relation of some triples can reduce the communication overhead, it will introduce computational overhead in some degree. Specifically speaking, according to the compression principles introduced above, the computational overhead of probability calculation without prior knowledge and conditional probability calculation is different, and as the number of compression rounds increases, the probability matrix that needs to be calculated in each round will increase nonlinearly. However, the triples that can be degraded in each round may not increase proportionally. 

As a consequence, defining a reasonable parameter that can describe the number of degenerate triples and can also connect computation and communication overhead is crucial, and the parameter is semantic compression ratio (SCR). If denote $u_k = \text{card} (\mathcal{U}_k), \forall k \in \mathcal{K} \cup \{0\}$, where $\mathcal{K} = \{1,2,\dots,K\}$, the SCR can be represented as
\begin{equation}
    \begin{aligned} 
         \Omega_k &= \dfrac{\sum_{j=1}^Q d_{kj}}{ \sum_{k=0}^K u_k},
    \end{aligned}
\label{SCR}
\end{equation}
where $d_{kj}$ denotes the number of the degenerate triples in the $j$-th round of compression of user $k$, and $k=0$ stands for the shared information.
\begin{figure}[t]
\centering
\includegraphics[width=0.8\linewidth]{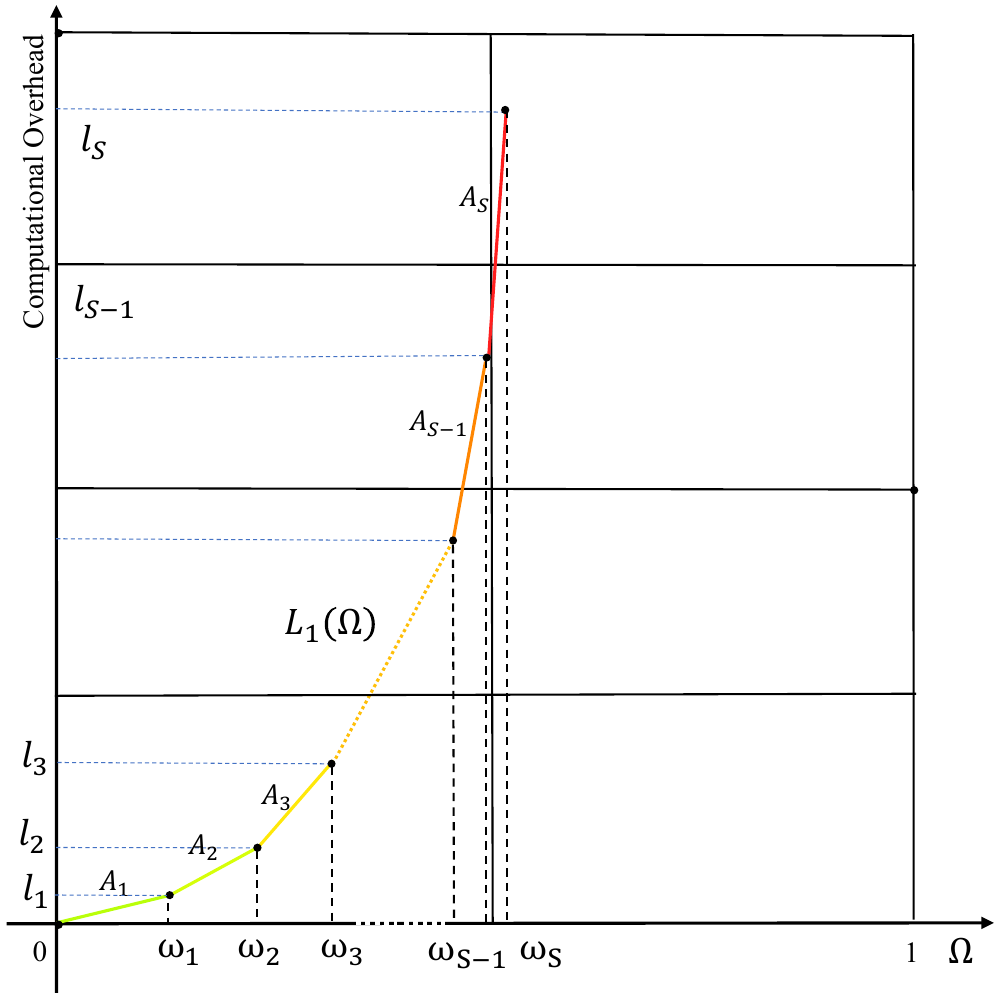}
\caption{Mapping relation between $\Omega$ and $L_1(\Omega)$} 
\label{piecewiseFunction}
\end{figure}

On the basis of the characteristic of the semantic information compression, we can fit the computational overhead $L_1(\Omega)$ using the piecewise function of the colored line in Fig.~\ref{piecewiseFunction}, and the segmented interval is related to the number of compression rounds. By local training, the values of compression ratio and computational overhead at the segmentation points can be obtained, thereby obtaining the slope and intercept of each line segment. Computational overhead is
\begin{equation}
    \begin{aligned} 
         L_1(\Omega_k) = A_s\Omega_k + B_s .\\
    \end{aligned}
\label{ComputationalOverhead}
\end{equation}
where $A_s$ and $B_s$ are the determined slope and intercept, respectively, with given SCR $\Omega_k$.

For communication overhead, equal-length coding is used to encode triples. Given the fact that the relation of a triple contains a larger amount of information, for $\varepsilon_i^j$, half of the code word is used to encode $r_i^j$, and the remaining half of the code word is used to encode $h_i$ and $t_i$. For convenience of description, denote that $h_i$ and $t_i$ both contain $R$ bits, and $r_i^j$ contains $2R$. Since the makers can be represented with very short bits, it can be omitted during the transmission. Communication overhead $L_2(\Omega)$ can be expressed as:
\begin{equation}
    \begin{aligned} 
         L_2(\Omega) = 2R(2-\Omega) ,\Omega\in[0,1].\\
    \end{aligned}
\label{CommunicationOverhead}
\end{equation}

After the compression process, we can obtain the output of semantic information compression model $\mathcal{C}$ as shown in Fig.~\ref{System Model Design}.

\subsection{Downlink RSMA Semantic Communication Model} 
Considering the feature that the triples are divided into shared ones and private ones, we utilize RSMA for the semantic information transmission of the multiple users. 

Signal $\mathbf{x}$, to be transmitted in the BS, can be written as:
\begin{equation}
    \begin{aligned} 
         \mathbf{x} = \sqrt{p_0}\mathbf{w_0}m_0 + \sum_{k=1}^{K} \sqrt{p_k}{\mathbf{w}}_km_k ,\\
    \end{aligned}
\label{x}
\end{equation}
where ${\mathbf{w}}_0$ is the transmit beamforming of the shared information $m_0$ consisting of the shared triples, ${\mathbf{w}}_k$ is the transmit beamforming of the private information $m_k$ consisting of the private triples of user $k$, $p_0$ stands for the transmit power allocated to shared information, and $p_k$ stands for the transmit power allocated to private information of user $k$.

For user $k$, the received information can be written as:
\begin{equation}
    \begin{aligned} 
         y_k = \mathbf{h}_k^H\mathbf{x} + n_k ,\\
    \end{aligned}
\label{y}
\end{equation}
where $\mathbf{h}_k$ stands for the channel gain between the BS and user $k$, and $n_k$ is the Gaussian noise with the power $\sigma^2$. Then the rate of user $k$ to decode the shared information is
\begin{equation}
    \begin{aligned} 
         s_k = B{\log}_2\left(1+\dfrac{p_0 |\mathbf{h}_k^H \mathbf{w}_0|^2}{\sum_{j=1}^K p_j |\mathbf{h}_k^H \mathbf{w}_j|^2+\sigma^2}\right) ,\\
    \end{aligned}
\label{SharedMessageRate}
\end{equation}
where $B$ is the bandwidth of BS. To ensure all the users can decode the shared information successfully, the rate of the shared information is supposed to be set as\cite{mao2018rate}:
\begin{equation}
    \begin{aligned} 
         s_0 = \mathop{\min}_{k\in \mathcal{K}}s_k .\\
    \end{aligned}
\label{so}
\end{equation}

Note that in addition to shared information and private information of user $k$, other messages are interference to user $k$, the rate of user $k$ to decode the private information can be presented as:
\begin{equation}
    \begin{aligned} 
         q_k = B{\log}_2\left(1+\dfrac{p_k |\mathbf{h}_k^H \mathbf{w}_k|^2}{\sum_{j=1,j\neq k}^K p_j |\mathbf{h}_k^H \mathbf{w}_j|^2+\sigma^2}\right) .\\
    \end{aligned}
\label{PrivateMessageRate}
\end{equation}
 
According to the coding rules of RSMA strategy\cite{9831440}, at receiver side, users first decode the received messages with successive interference cancellation (SIC) technique and obtain the received semantic information $\mathcal{C'}$. 

With the help of artificial intelligence generated content (AIGC) model employed in user side, the received semantic information $\mathcal{C'}$ can generate the source data output $\mathcal{D'}$.

\section{Problem Formulation and Algorithm Design} \label{Algorithm Design}
In this section, we first formulate the problem on minimizing the system energy consumption comprehensively considering the communication and computation during transmission. Then, an alternative algorithm is proposed by optimizing sub-problems of joint power allocation and beamforming design, computation capacity allocation and semantic compression ratio. The joint power allocation and beamforming design problem can be approximated by a convex problem and then solved by existing convex optimization toolbox. Computation capacity allocation sub-problem can be proven to be convex after some processing. Semantic compression ratio sub-problem can be derived to obtain its optimal closed-form solution.

\subsection{Problem Formulation}
For a transmission task involving $K$ users and one BS, assume that there are $u_0$ shared triples and $u_k$ private triples of user $k$. After certain rounds of compression, assume the SCR of shared triples is $\Omega_0$ and the SCR of private triples of user $k$ is $\Omega_{k}$, which consist the compression vector $\mathbf{\Omega} = [\Omega_0, \Omega_1, \dots, \Omega_K]^T$.

Define $\mathbf{f} = [f_0,f_1,\dots,f_K]^T$, where $f_k$ is the computation capacity allocated for user $k$, and $f_0$ is the computation capacity allocated for compressing shared information. For user $k$, jointly consider the compression for the semantic information in the BS and transmission from the BS to user $k$, and the time delay can be presented as:
\begin{equation}
    \begin{aligned} 
         \tau_k = \dfrac{L_1(\Omega_{k})}{f_k} + \dfrac{u_{k}L_2(\Omega_{k})}{q_k}.\\
    \end{aligned}
\label{TimedelayPrivate}
\end{equation}

For shared information, its time delay can be described as:
\begin{equation}
    \begin{aligned} 
         \tau_0 = \dfrac{L_1(\Omega_{0})}{f_0} + \dfrac{u_0L_2(\Omega_{0})}{s_0}.\\
    \end{aligned}
\label{TimedelayShared}
\end{equation}

In order to ensure that semantic information has a satisfying accuracy between the transmitter and the receiver, the semantic accuracy parameter $A(\mathcal{S},\mathcal{S}')$ is introduced\cite{wang2022performance}:
\begin{equation}
    \begin{aligned} 
         A_k(\mathcal{S}_k,\mathcal{S}'_k) = \dfrac{\sum_{i=1}^I {\min}\left\{\sigma(\mathcal{S}_k,\varepsilon_i),\sigma(\mathcal{S}'_k,\varepsilon_i)\right\}}{\sum_{i=1}^I {\min}\left\{\sigma(\mathcal{S}_k,\varepsilon_i)\right\}},\\
    \end{aligned}
\label{SemanticAccuracy}
\end{equation}
where $I$ means the total number of the different triples in the semantic information set $\mathcal{S}_k$ is $I$, $\mathcal{S}'_k$ is the semantic information received and recovered by user $k$, and $\sigma(\mathcal{S},\varepsilon_i)$ is the number of occurrences of $\varepsilon_i$ in $\mathcal{S}$.

The total computation energy consumption includes the energy consumption caused by the compression of both shared information and each user's private information, i.e.,
\begin{equation}
    \begin{aligned} 
         E_1 = \sum_{k=0}^K \xi L_1(\Omega_k){f_k}^2,\\
    \end{aligned}
\label{ComputationEnergy}
\end{equation}
where $\xi$ is a constant coefficient to measure the effective
switched capacitance.

For communication part, the total communication energy consumption have both shared part and private part. If define $\mathbf{p} = [p_0,p_1,\dots,p_K]^T$, where $p_0$ is the power allocated to sent the shared information, $p_k$ is the power allocated to sent the private information of user $k$, total communication energy consumption is
\begin{equation}
    \begin{aligned} 
         E_2 = \dfrac{u_0 L_2(\Omega_0)}{s_o} + \sum_{k=1}^K \dfrac{u_kL_2(\Omega_k)}{q_k}.\\
    \end{aligned}
\label{communicationEnergy}
\end{equation}

If assume $\mathbf{W} = [\mathbf{w_0},\mathbf{w_1},\dots,\mathbf{w_K}] $ is the matrix representing the beamforming of the users, where $\mathbf{w}_k$ is the beamforming vector of user $k$, based on \eqref{ComputationEnergy} and \eqref{communicationEnergy}, we can construct the following joint optimization problem with the objective function that is to minimize the total energy consumption of the system.
\begin{subequations}\label{OptimizationProblem}
    \begin{align} 
        \mathop{\min}_{\mathbf{p}, \mathbf{W},\mathbf{f},\mathbf{\Omega}} \quad & E = E_1+E_2 ,\tag{\ref{OptimizationProblem}}\\
         \textrm{s.t.} \qquad 
         & \sum_{k=0}^K p_k \leq P_{\text{max}},\\
         & \sum_{k=0}^K f_k \leq F_{\text{max}},\\
         & \tau_k \leq T_{\text{max}}, \forall k \in \mathcal{K} \cup \{0\},\\
         & A_k(\{S\},\{S'_k\}) \geq A_{\text{min}}, \forall k \in \mathcal{K} \cup \{0\}, \\
         & f_k \geq 0, \forall k \in \mathcal{K} \cup \{0\},\\
         & p_k \geq 0, \forall k \in \mathcal{K} \cup \{0\},\\
         & 0 \leq \Omega_k \leq 1,\forall k \in \mathcal{K} \cup \{0\},\\
         & \|\mathbf{w}_k\| = 1, \forall k \in \mathcal{K} \cup \{0\},
    \end{align}
\end{subequations}
where $P_{\text{max}}$ is the total maximum transmission power, $F_{\text{max}}$ is the maximum computation capacity, $T_{\text{max}}$ is the maximum time delay acceptable to all the users and $A_{\text{min}}$ is the minimum value of the semantic accuracy. The objective of the optimization problem is to minimize the energy consumption of the semantic communication system while obeying all constraints.

Optimizing problem \eqref{OptimizationProblem} involves many variables, and the objective function and constraints cannot be directly converted into convex. Therefore, we propose an alternating optimization method to solve the problem, which is decomposed the original problem into the following three sub-problems. In the optimization in the following three subsection, we omit the constant in the objective function.

\subsection{Joint Optimization of Power Allocation and Beamforming Design}
When given the semantic compression ratio vector and computation capacity vector, the constraints (\ref{OptimizationProblem}b), (\ref{OptimizationProblem}e), (\ref{OptimizationProblem}g) and (\ref{OptimizationProblem}h) can be taken away from consideration. The sub-problem of jointly optimize power control and beamforming design can be formulated as:
\begin{subequations}\label{SubOptimizationProblem_Power}
    \begin{align} 
    \mathop{\min}_{\mathbf{p}, \mathbf{W}} \;&  
		 E = \xi \sum_{k=0}^K \big(A_k \Omega_k + B_k\big){f_k}^2+2R\Big[\dfrac{u_0\big(2-\Omega_0\big)p_{0}}{s_0}
   \nonumber \\
   &+\sum_{j=1}^K \dfrac{u_j\big(2-\Omega_j\big)p_j}{q_j}\Big], \tag{\ref{SubOptimizationProblem_Power}}\\
         \textrm{s.t.} \qquad & \sum_{k=0}^K p_k \leq P_{\text{max}},\\
         & \dfrac{2Ru_k(2-\Omega_k)}{q_k} + \dfrac{A_k \Omega_k + B_k}{f_k} \leq T_{\text{max}}, \forall k \in \mathcal{K},\\ 
         & \dfrac{2Ru_0(2-\Omega_0)}{s_0} + \dfrac{A_0 \Omega_0 + B_0}{f_0} \leq T_{\text{max}}\\
         & p_k \geq 0, \forall k \in \mathcal{K} \cup \{0\},\\
         & \|\mathbf{w}_k\| = 1, \forall k \in \mathcal{K} \cup \{0\}.
    \end{align}
\end{subequations}

Due to the fact that both the objective function and the constraints (\ref{SubOptimizationProblem_Power}b) and (\ref{SubOptimizationProblem_Power}e) are all non-convex, problem (\ref{SubOptimizationProblem_Power}) is non-convex. To handle this, first and foremost, it is obvious that when $\mathbf{\Omega}$ and $\mathbf{f}$ are given, the first and the last terms of the objective function in (\ref{SubOptimizationProblem_Power}) are constants. Then to figure the non-convexity of the middle term, i.e. the energy consumption caused by communication, we utilize the parameter ${p_k}^2 $ to replace $p_k$ for every $k \in \mathcal{K} \cup \{0\}$ and introduce the slack parameter $\bm{\zeta} = [\zeta_0,\zeta_1,\dots,\zeta_K]^T$, and then problem (\ref{SubOptimizationProblem_Power}) can be equivalently transformed to:
\begin{subequations}\label{SubOptimizationProblem_Power_new}
    \begin{align} 
    \mathop{\min}_{\mathbf{p}, \mathbf{W}, \bm{\zeta}} \;&  
		 2R\Big[\dfrac{u_0\big(2-\Omega_0\big)p^2_{0}}{\zeta_0} +\sum_{j=1}^K \dfrac{u_j\big(2-\Omega_j\big)p^2_j}{\zeta_j}\Big], \tag{\ref{SubOptimizationProblem_Power_new}}\\
         \textrm{s.t.} \qquad & \sum_{k=0}^K p_k^2 \leq P_{\text{max}},\\
         & \dfrac{2Ru_k(2-\Omega_k)}{\zeta_k} \leq {T_k}', \forall k \in \mathcal{K}, \\
         & \dfrac{2Ru_0(2-\Omega_0)}{\zeta_0} \leq T_0',\\
         \nonumber
         & \zeta_0 \leq B{\log}_2\left(1+\dfrac{p_0^2 |\mathbf{h}_k^H \mathbf{w}_0|^2}{\sum_{j=1}^K p_j^2 |\mathbf{h}_k^H \mathbf{w}_j|^2+\sigma^2}\right),\\ &\forall k \in \mathcal{K},\\
         \nonumber
         & \zeta_k \leq B{\log}_2\left(1+\dfrac{p_k^2 |\mathbf{h}_k^H \mathbf{w}_k|^2}{\sum_{j=1,j\neq k}^K p_j^2 |\mathbf{h}_k^H  
         \mathbf{w}_j|^2+\sigma^2}\right),\\
         & \forall k \in \mathcal{K},\\
         & \zeta_k, p_k \geq 0, \forall k \in \mathcal{K} \cup \{0\},\\
         & \|\mathbf{w}_k\| \leq 1, \forall k \in \mathcal{K} \cup \{0\}.
    \end{align}
\end{subequations}
where constraint (\ref{SubOptimizationProblem_Power}e) is replaced by inequality without loss of generality an $T_k'$ are constants,
\begin{align} 
        {T_k}' &= T_{\text{max}} - \dfrac{A_k \Omega_k + B_k}{f_k}, \forall k \in \mathcal{K} \cup \{0\}.
    \end{align}
\label{C1andT'}

The reason why $p_k^2$ is introduced to replace the $p_k$ is that the Hessian matrix of function $\dfrac{{p_k}^2}{\zeta_k}$ is semi-positive definite,
\begin{align}\label{Hessian}
\mathbf{H}_k=
\left [ \begin{matrix} 
    \frac{\partial^2 \frac{p_k^2}{\zeta_k}}{\partial p_k^2} &  \frac{\partial^2 \frac{p_k^2}{\zeta_k}}{\partial p_k\partial \zeta_k}  \\
  \frac{\partial^2 \frac{p_k^2}{\zeta_k}}{\partial \zeta_k\partial \zeta_k}    & \frac{\partial^2\frac{p_k^2}{\zeta_k}}{\partial \zeta_k^2} 
\end{matrix} \right ]
= \frac{2}{\zeta_k^3} 
\left [ \begin{matrix} 
    \zeta_k^2 & -p_k\zeta_k  \\
  -p_k\zeta_k&  p_k^2
\end{matrix} \right ],
\end{align}
whose quadratic form is:
\begin{align}\label{Hessian_quadratic}
\mathbf{v}^H\mathbf{H}_k\mathbf{v}=
(v_1\zeta_k-v_2p_k)^2
\geq 0,
\end{align}
where $\mathbf{v} = [v_1, v_2]^T$ stands for an arbitrary non-zero vector. Therefore function $\dfrac{{p_k}^2}{\zeta_k}$ is convex. Consequently, the objective function is in the form of the sum of ($K+1$) convex functions and a constant, i.e. it is convex. Subsequently, we need to tackle the non-convexity of constraint (\ref{SubOptimizationProblem_Power_new}d) and (\ref{SubOptimizationProblem_Power_new}e). 

Through Introducing slack variable  $\bm{\alpha} = [\alpha_1,\dots,\alpha_K]^T$ and $\bm{\beta} = [\beta_1,\dots, \beta_K]^T$, constraint (\ref{SubOptimizationProblem_Power_new}) can be reformulated as:
\begin{subequations}\label{SubOptimizationProblem_Power_new2}
    \begin{align} 
    \mathop{\min}_{\mathbf{p}, \mathbf{W}, \bm{\zeta}, \bm{\alpha}, \bm{\beta}} \;&  
		 2R\Big[\dfrac{u_0\big(2-\Omega_0\big)p^2_{0}}{\zeta_0} +\sum_{j=1}^K \dfrac{u_j\big(2-\Omega_j\big)p^2_j}{\zeta_j}\Big], \tag{\ref{SubOptimizationProblem_Power_new2}}\\
         \textrm{s.t.} \qquad & \sum_{i=0}^K p_i^2 \leq P_{\text{max}},\\
         & \dfrac{2Ru_k(2-\Omega_i)}{\zeta_k} , \leq {T_k}', \forall k \in \mathcal{K}, \\
         & \dfrac{2Ru_0(2-\Omega_0)}{\zeta_0} \leq T_0'\\
         & \zeta_0 \leq B{\log}_2\left(1+\alpha_k\right), \forall k \in \mathcal{K},\\
         & \zeta_k \leq B{\log}_2\left(1+\beta_k \right), \forall k \in \mathcal{K},\\
         & \|\mathbf{w}_k\| \leq 1, \forall k \in \mathcal{K} \cup \{0\},\\
         & \zeta_k, p_k \geq 0, \forall k \in \mathcal{K} \cup \{0\},\\
         & \alpha_k \leq \dfrac{p_0^2 |\mathbf{h}_k^H \mathbf{w}_0|^2}{\sum_{j=1}^K p_j^2 |\mathbf{h}_k^H \mathbf{w}_k|^2+\sigma^2}, \forall k \in \mathcal{K},\\
         & \beta_k \leq \dfrac{p_k^2 |\mathbf{h}_k^H \mathbf{w}_k|^2}{\sum_{j=1,j\neq k}^K p_j^2 |\mathbf{h}_k^H  
         \mathbf{w}_j|^2+\sigma^2}, \forall k \in \mathcal{K}.
    \end{align}
\end{subequations}

In problem (\ref{SubOptimizationProblem_Power_new2}), constraints (\ref{SubOptimizationProblem_Power_new2}h) and (\ref{SubOptimizationProblem_Power_new2}i) are non-convex, and we utilize successive convex approximation (SCA) method to tackle this.

Introduce slack variable $\bm{\gamma} = [\gamma_1,\dots,\gamma_K]^T$, and constraint (\ref{SubOptimizationProblem_Power_new2}h) can be equivalent to
\begin{equation}
    \begin{aligned} 
         \sum_{j=1}^K p_j^2 |\mathbf{h}_k^H  
         \mathbf{w}_j|^2+\sigma^2 \leq \gamma_k, \forall k \in \mathcal{K},\\
    \end{aligned}
\label{InterferPowerforShared}
\end{equation}
\begin{equation}
    \begin{aligned} 
         \alpha_k \gamma_k \leq p_0^2 |\mathbf{h}_k^H \mathbf{w}_0|^2, \forall k \in \mathcal{K},\\
    \end{aligned}
\label{SignalPowerforShared}
\end{equation}

For constraint (\ref{InterferPowerforShared}), the left hand of it can be written as:
\begin{align}
         & \sum_{j=1}^K \dfrac{1}{4} \Big[\left(p_j^2 + |\mathbf{h}_k^H \mathbf{w}_j|^2\right)^2 - (p_j^2 - |\mathbf{h}_k^H \mathbf{w}_j|^2)^2\Big] + \sigma^2.
         \label{InterferPowerforShared_new}
\end{align}

To tackle the non-convexity of (\ref{InterferPowerforShared_new}), we replace it with its first-order Taylor approximation and we can get
\begin{align}
    \nonumber
    &\sum_{j=1}^K \dfrac{1}{4} \Bigg\{ \left(p_j^2 + |\mathbf{h}_k^H \mathbf{w}_j|^2\right)^2  - \left[\left(p_j^{(n)}\right)^2 - |\mathbf{h}_k^H \mathbf{w}^{(n)}_j|^2 \right]^2\\
    \nonumber
    &+4\left[\left(p_j^{(n)}\right)^2 - |\mathbf{h}_k^H\mathbf{w}_j^{(n)}|^2\right]  p_j^{(n)}\left(p_j-p_j^{(n)}\right) + 4\left[\left(p_j^{(n)}\right)^2 \right.\\
    \nonumber
    & - |\mathbf{h}_k^H\mathbf{w}_j^{(n)}|^2\Bigg] \left({Re}[\mathbf{h}_k^H \mathbf{w}_j^{(n)}\mathbf{w}_j^H\mathbf{h}_i] -
    |\mathbf{h}_k^H \mathbf{w}_j^{(n)}|^2\right)\Bigg\}\\
    & +\sigma^2  \leq \gamma_k , \forall k \in \mathcal{K},
    \label{InterferPowerforShared_new_Taloy}
\end{align}
where the superscript $(n)$ means the value of the variable in the $n$-th iteration.

The same story, for (\ref{SignalPowerforShared}), it can be written as
\begin{align}
    \nonumber
    &\dfrac{1}{4} \left[ \left( \alpha_k + \gamma_k\right)^2 - \left( \alpha_k - \gamma_k\right)^2 \right] \leq \dfrac{1}{4} \Big[\big(p_0^2 + |\mathbf{h}_k^H\mathbf{w}_0|^2\big)^2 \\
    &  - \big(p_0^2 - |\mathbf{h}_k^H\mathbf{w}_0|^2\big)^2\Big], \forall k \in \mathcal{K}. 
         \label{SignalPowerforShared_new}
\end{align}

It can be acknowledged that we cannot make $\mathbf{h}_k^H\mathbf{w}_0$ always a real value for arbitrary $k \in \mathcal{K}$ at the same time by changing the beamforming $\mathbf{w}_0$, hence we replace both side of (\ref{SignalPowerforShared_new}) with respectively first-order Taylor approximation. For $\forall k \in \mathcal{K}$,
\begin{align}
    \nonumber
     &\left( \alpha_k + \gamma_k\right)^2 - \left( \alpha_k^{(n)} - \gamma_k^{(n)}\right)^2 - 2\left( \alpha_k^{(n)} - \gamma_k^{(n)}\right)(\alpha_k-\alpha_k^{(n)}) \\
    \nonumber
    &+ 2\left( \alpha_k^{(n)} - \gamma_k^{(n)}\right)(\gamma_k-\gamma_k^{(n)})   \\
    \nonumber
    & \leq  \left[\left(p_0^{(n)}\right)^2 + |\mathbf{h}_k^H \mathbf{w}_0^{(n)}|^2\right]^2- \left(p_0^2 - |\mathbf{h}_k^H \mathbf{w}_0|^2 \right) ^2 \\
    \nonumber
    & + 4\left[\left(p_0^{(n)}\right)^2 + |\mathbf{h}_k^H \mathbf{w}_0^{(n)}|^2 \right] p_0^{(n)}\left(p_0-p_0^{(n)}\right) + 4\left[\left(p_0^{(n)}\right)^2 \right. \\
    & + |\mathbf{h}_k^H \mathbf{w}_0^{(n)}|^2 \Big] {Re}\Big[\mathbf{h}_k^H \mathbf{w}_0^{(n)}\mathbf{w}_0^H\mathbf{h}_k -|\mathbf{h}_k^H \mathbf{w}_0^{(n)}|^2\Big] +\sigma^2,
         \label{SignalPowerforShared_new_Taylor}
\end{align}
which is convex now. 

For constraint (\ref{SubOptimizationProblem_Power_new2}i), without loss of generality, it can also be equivalent to
\begin{equation}
    \begin{aligned} 
         \sum_{j=1,j\neq k}^K p_j^2 |\mathbf{h}_k^H  
         \mathbf{w}_j|^2+\sigma^2 \leq \eta_k, \forall k \in \mathcal{K},\\
    \end{aligned}
\label{InterferPower}
\end{equation}
\begin{equation}
    \begin{aligned} 
         \beta_k \eta_k \leq p_k^2 |\mathbf{h}_k^H \mathbf{w}_k|^2, \forall k \in \mathcal{K},\\
    \end{aligned}
\label{SignalPower}
\end{equation}
where $\eta_k$ is a non-negative slack variable. 

For constraint (\ref{InterferPower}), the left hand of it can be written as:
\begin{align}
         \sum_{j=1,j\neq k}^K \dfrac{\left(p_j^2 + |\mathbf{h}_k^H \mathbf{w}_j|^2\right)^2  - (p_j^2 - |\mathbf{h}_k^H \mathbf{w}_j|^2)^2}{4}  + \sigma^2.
         \label{InterferPower_new}
\end{align}

Replace (\ref{InterferPower_new}) with its first-order Taylor approximation and we can get the new constraint:
\begin{align}
    \nonumber
    &\sum_{j=1,j\neq k}^K \dfrac{1}{4} \Bigg\{ \left(p_j^2 + |\mathbf{h}_k^H \mathbf{w}_j|^2\right)^2 - \left[\left(p_j^{(n)}\right)^2 - |\mathbf{h}_k^H \mathbf{w}_j^{(n)}|^2\right]^2\\
    \nonumber
    & - 4\left[\left(p_j^{(n)}\right)^2 - |\mathbf{h}_k^H\mathbf{w}_j^{(n)}|^2\right]  p_j^{(n)}\left(p_j-p_j^{(n)}\right) + 4\left[\left(p_j^{(n)}\right)^2 \right.\\
    \nonumber
    & - |\mathbf{h}_k^H\mathbf{w}_j^{(n)}|^2\Bigg] \left({Re}[\mathbf{h}_k^H \mathbf{w}_j^{(n)}\mathbf{w}_j^H\mathbf{h}_k] -
    |\mathbf{h}_k^H \mathbf{w}_j^{(n)}|^2\right)\Bigg\}\\
    &+\sigma^2  \leq \eta_k, \forall k \in \mathcal{K},
\label{InterferPower_new_Taloy}
\end{align}

 Moreover, by changing the phase of beamforming, we can always get $\mathbf{h}_k^H\mathbf{w}_k$ in (\ref{SignalPower}) as a real value. Hence, (\ref{SignalPower}) can be rewritten as:
\begin{equation}
    \begin{aligned}  
     \dfrac{\sqrt{\beta_k \eta_k}}{p_k} \leq \mathbf{h}_k^H \mathbf{w}_k, \forall k \in \mathcal{K}.\\
    \end{aligned}
\label{SignalPower_new}
\end{equation}

Faced with the fact that the right hand of (\ref{SignalPower_new}) is linear, we are going to use the first-order Taylor series to expand the non-convex left hand side of it. Then constraint (\ref{SignalPower_new}) can be rewritten as:

    \begin{align} 
    \nonumber
         &\dfrac{\sqrt{\beta_k^{(n)}\eta_k^{(n)}}}{p_k^{(n)}} + \dfrac{\eta_k^{(n)}}{p_k^{(n)}\sqrt{\beta_k^{(n)}}}(\beta_k-\beta_k^{(n)}) + \dfrac{\beta_k^{(n)}}{p_k^{(n)}\sqrt{\eta_k^{(n)}}}(\eta_k-\eta_k^{(n)})\\ 
         &-\dfrac{\sqrt{\beta_k^{(n)}\eta_k^{(n)}}}{(p_k^{(n)})^2}(p_k-p_k^{(n)}) \leq \mathbf{h}_k^H \mathbf{w}_k, \forall k \in \mathcal{K}
         \label{SignalPower_new_Taylor}
    \end{align}

After operating Taylor approximations, we obtain the convex approximations of (\ref{InterferPowerforShared_new_Taloy}) (\ref{SignalPowerforShared_new_Taylor}) and (\ref{InterferPower_new_Taloy}) (\ref{SignalPower_new_Taylor}) which are respectively equivalent to the constraint (\ref{SubOptimizationProblem_Power_new2}g) and (\ref{SubOptimizationProblem_Power_new2}h). Therefore problem (\ref{SubOptimizationProblem_Power_new2}) can be further approximated as
\begin{subequations}\label{SubOptimizationProblem_Power_new3}
    \begin{align} 
    \mathop{\min}_{\mathbf{p}, \mathbf{W}, \bm{\zeta}, \bm{\alpha}, \bm{\beta}, \bm{\gamma}, \bm{\eta}} \;&  
		 2R\Big[\dfrac{u_0\big(2-\Omega_0\big)p^2_{0}}{\zeta_0} +\sum_{j=1}^K \dfrac{u_j\big(2-\Omega_j\big)p^2_j}{\zeta_j}\Big], \tag{\ref{SubOptimizationProblem_Power_new3}}\\
         \textrm{s.t.} \qquad 
         & (\ref{SubOptimizationProblem_Power_new2}a) - (\ref{SubOptimizationProblem_Power_new2}g),
         (\ref{InterferPowerforShared_new_Taloy}), (\ref{SignalPowerforShared_new_Taylor}), (\ref{InterferPower_new_Taloy}), (\ref{SignalPower_new_Taylor}),\\
         &  \eta_k, \gamma_k \geq 0,\forall k \in \mathcal{K},
    \end{align}
\end{subequations}
which can be solved by existing optimization tool box.

\subsection{Semantic Compression Ratio Optimization}
There are two difficulties standing in the way concerning compression ratio, i.e. the constraint of implicit expression of semantic accuracy, which has no mathematical expressions and the computational overhead function isn't smooth.

To deal with first difficulty, we first analyze the trend of accuracy and compression ratio. Considering the fact that more correct information is used to recover the semantic information, the higher semantic accuracy can be gotten, the semantic accuracy always decreases with the compression ratio increases. Thus, constraint (\ref{OptimizationProblem}a) can be equivalent to
\begin{align}  
       \Omega_k \leq \Theta,
        \label{SemanticAccuracyMAX}
\end{align}
where $\Theta$ is the compression ratio corresponding to minimum semantic accuracy $A_{\text{min}}$ and can be obtained via simulations.

To address the second difficult, we  first review Fig.\ref{piecewiseFunction}. The slopes of  $\mathop{L_1(\Omega)}$ do not decrease, so $\mathop{L_1(\Omega)}$ is convex, i.e., 
\begin{align}  
       L_1(\Omega_k) = \max\limits_{i=1,2\dots,S}(A_i\Omega_k + B_i).
        \label{convexPieceFunction}
\end{align}

When given the the power allocation, beamforming design, and computation allocation, the compression rate optimization sub-problem can be written as
\begin{subequations}\label{SubOptimizationProblem_Compression}
    \begin{align} 
        \mathop{\min}_{\mathbf{\Omega}} \quad 
         & E = \xi \sum_{k=0}^K\Big[\max\limits_{i=1,2\dots,S}(A_i\Omega_k + B_i)\Big]f_k^2\nonumber\\
         & -2R(\dfrac{u_0\Omega_0p_0}{s_0}+\sum_{j=1}^K\dfrac{u_j\Omega_j p_j}{q_j}) ,\tag{\ref{SubOptimizationProblem_Compression}}\\
         \textrm{s.t.} \qquad 
         \nonumber
         & \dfrac{\max\limits_{i=1,2\dots,S}(A_i\Omega_k + B_i)}{f_i}+2R\dfrac{(2-\Omega_k)u_k}{q_k}\} \leq T_{\text{max}}, \\
         & \forall k \in \mathcal{K}\\
         & \dfrac{ \max\limits_{i=1,2\dots,S}(A_i\Omega_0 + B_i)}{f_0} +2R\dfrac{(2-\Omega_0)u
         _0}{s_0}  \leq T_{\text{max}},\\
         & 0 \leq \Omega_k \leq \Theta, \forall k \in \mathcal{K} \cup \{0\}.
    \end{align}
\end{subequations}

It can be observed that the optimization objective in problem (\ref{SubOptimizationProblem_Compression}) is to sum $K+1$ convex functions and subtract $K+1$ linear functions, i.e. the objective function is convex, and its constraints are obvious convex constraints. Therefore, existing optimization tools can be utilized to solve the problem.

\subsection{Computation Capacity Allocation Optimization}
When given semantic compression ratio, power allcation vector and beamforming design, the sub-problem of
computation capacity can be formulated as:
\begin{subequations}\label{SubOptimizationProblem_Computation}
    \begin{align} 
        \mathop{\min}_{\mathbf{f}} \quad 
         & E = \xi \sum_{k=0}^K(A_k\Omega_k+B_k)f_k^2,\tag{\ref{SubOptimizationProblem_Computation}}\\
         \textrm{s.t.} \qquad 
         & \sum_{k=0}^K f_k \leq F_{\text{max}},\\
         & \dfrac{ (A_k \Omega_k + B_k)}{f_k} \leq {T_k}^{''}, \forall k \in \mathcal{K}\cup \{0\},\\
         & f_k \geq 0, \forall k \in \mathcal{K} \cup \{0\},
    \end{align}
\end{subequations}
where
\begin{align} 
        {T_i}'' &= T_{\text{max}} - \dfrac{2Ru_k(2-\Omega_k)}{q_k},\forall k \in \mathcal{K}.\\
        {T_0}'' &= T_{\text{max}} - \dfrac{2Ru_0(2-\Omega_0)}{s_0}.
     \label{C2andT''}
\end{align}

Considering the fact that time delay caused by computation cannot be zero, i.e. $T'' \neq 0$, constraint (\ref{SubOptimizationProblem_Computation}b) is equivalent to
\begin{equation}
    \begin{aligned} 
        f_k \geq \dfrac{A_k\Omega_k + B_k}{{T_k}''} > 0, \forall k \in \mathcal{K} \cup \{0\}.
    \end{aligned}
\label{fiNonnegative}
\end{equation}

Construct Lagrange function of problem (\ref{SubOptimizationProblem_Computation}) with Lagrange multiplier $\lambda_1$ and $\bm{\mu} = [\mu_0,\mu_1\dots,\mu_K]^T$, and we can get
\begin{align} 
        \nonumber  
        &\mathcal{L}(\mathbf{f},\lambda_1,\bm{\mu}) = \xi \sum_{k=0}^K(A_k\Omega_k+B_k)f_k^2 \\
        & + \lambda_1(\sum_{i=0}^K f_k - F_{\text{max}}) + \sum_{k=0}^K\mu_k \left( \dfrac{A_k\Omega_k + B_k}{{T_k}''} - f_k \right),
        \label{LagrangeFunction_Computation}
\end{align}
where $\lambda_1 \geq 0$ and $\mu_k \geq 0$. We can first get the derivation of (\ref{LagrangeFunction_Computation}) about $f_k$ and make it equal to zero, then we can obtain
\begin{align}
        &\dfrac{\partial \mathcal{L}(\mathbf{f} ,{\lambda}_{1} , \bm{\mu})}{\partial f_k} = 2 \xi (A_k\Omega_k +B_k)f_k + \lambda_1 - \mu_k = 0,
        \label{LagrangeFunction_Computation_Derivation}
\end{align}
i.e.,
\begin{align}  
        f_k = \dfrac{\mu_k-\lambda_1}{\xi(A_k\Omega_k +B_k)}, \forall k \in \mathcal{K} \cup \{0\}
        \label{Computation_optimal}
\end{align}

From (\ref{fiNonnegative}), we know that any $f_k$ is non-negative, thus numerator of (\ref{Computation_optimal}) is non-negative. Because $\lambda_1 \geq 0$, $u_k > 0$. According to complementary slackness of (\ref{LagrangeFunction_Computation}), we can get the closed-form solution to (\ref{SubOptimizationProblem_Computation})
\begin{align}  
       f_k^* = \dfrac{A_k\Omega_k + B_k}{{T_k}''} , \forall k \in \mathcal{K} \cup \{0\}.
        \label{Computation_optimal3}
\end{align}

\subsection{Algorithm Design and Analysis}
The overall green probabilistic semantic communication resource allocation is presented in Algorithm~\ref{Algorithm1}. The complexity of solving problem (\ref{OptimizationProblem}) is determined by the solving process of the three sub-problems at each iteration. According to \cite{lobo1998applications}, the complexity of acquiring the optimal solution to problem (\ref{SubOptimizationProblem_Power_new3}) is $\mathcal{O}(Y_1^2Y_2)$, where $Y_1 = (L+6)K + L +2$ is the number of the variables and $Y_2 = 12K + 5$ is the number of the constraints. Therefore, the total complexity of solving joint power allocation and beamforming design sub-problem is $\mathcal{O}(I_1L^2K^3)$, where $I_1$ is the number of iterations for the SCA method. The semantic compression ratio optimization sub-problem involves the complexity of $\mathcal{O}(K^3S)$ with the interior-point method for solving the convex problem. Given the fact that problem (\ref{SubOptimizationProblem_Computation}) has the closed-form solution, the complexity of it is $\mathcal{O}(K)$. Hence, the total complexity of solving problem (\ref{OptimizationProblem}) is $\mathcal{O}(I_1I_2L^2K^3 + I_2K^3S + I_2K) = \mathcal{O}(I_1I_2L^2K^3)$, where $I_2$ is the number of outer iterations of Algorithm~\ref{Algorithm1}.

 \begin{algorithm}[t]
 	\caption{Alternating Algorithm for Problem \eqref{OptimizationProblem}}
 	  \label{Algorithm1}
    \KwIn{Initialized parameters $\mathbf{p}^{(0)}$, $\mathbf{f}^{(0)}$, $\mathbf{W}^{(0)}$, $\mathbf{\bm{\Omega}^{(0)}}$; $i = 0$, and maximum iteration number $I_2$}
    \KwOut{Minimal system energy consumption $E_{\text{min}}$; Optimal parameters $\mathbf{p}^*$, $\mathbf{f}^*$, $\mathbf{W}^*$, $\bm{\Omega}^*$}
    \BlankLine
    \While{\textnormal{$i \neq I_2$}}{
    Solve problem \eqref{SubOptimizationProblem_Power} through solving a set of convex sub-problems \eqref{SubOptimizationProblem_Power_new3} using the successive convex approximation method with given $\mathbf{f}^{(i-1)}$ and $\bm{\Omega}^{(i-1)}$ and obtain the optimal solution for this sub-problem at this iteration $\mathbf{W}^{(i)}$ and $\mathbf{p}^{(i)}$.
    
    Solve problem \eqref{SubOptimizationProblem_Compression} using the existing optimize tool with given $\mathbf{f}^{(i-1)}$, $\mathbf{W}^{(i)}$ and $\mathbf{p}^{(i)}$, and obtain the optimal solution for this sub-problem at this iteration $\bm{\Omega}^{(i)}$.

     Solve problem \eqref{SubOptimizationProblem_Computation} via utilizing the derived closed-form of optimal solutions to it with given $\mathbf{\bm{\Omega}}^{(i)}$, $\mathbf{W}^{(i)}$ and $\mathbf{p}^{(i)}$, and obtain the optimal solution for this sub-problem at this iteration $\mathbf{f}^{(i)}$.

    Calculate the value of the objective function (\ref{OptimizationProblem}).
    }
 		
 \end{algorithm}

\section{Simulation Results and Analysis} \label{Simulation Results and Analysis}
In this section, we first accomplish the verification of the correctness of solving the three sub-problems via MATLAB simulations, and subsequently, based on the alternating optimization algorithm proposed in Algorithm~\ref{Algorithm1}, we completed the solution to problem (\ref{OptimizationProblem}) and obtained corresponding results.

\subsection{Sub-problems Solving Verification}
In simulations, we set the bandwidth of the BS is $B = 20 \rm{MHz}$ and the maximum transmit power $P_{\text{max}}$ and computation capacity $F_{\text{max}}$ are $30\rm{dBm}$ and $10^9\rm{Hz}$, respectively. The allowed maximum time delay is set as $T_{\text{max}} = 1\rm{s}$. The constants, effective switched capacitance $\xi = 10^{-28}$, the information contained in every triple $4R = 4\times32\rm{bit}$, and the power spectral density of the noise $\sigma^2 = -174\rm{dBm/Hz}$, are set as Table~\ref{systemParameter} shown.

\begin{table}[t]
\caption{MAIN SYSTEM PARAMETERS}
\centering
\begin{tabular}{c|c|c}
\hline
\textbf{Parameter}    & \textbf{Symbol} & \textbf{Value} \\ \hline
Bandwidth of the B    & $B$             & $20 \rm{MHz}$         \\
Maximum transmit power & $P_{\text{max}}$  & $30 \rm{dBm}$          \\
Effective switched capacitance                & $\xi$           & $10^{-28}$             \\
Power spectral density of the noise       & $\sigma^2$     & $-174 \rm{dBm/Hz}$        \\
Maximum time delay                  & $T_{\text{max}}$     & $1\rm{s}$         \\
Maximum computation capacity        & $F_{\text{max}}$      & $10^{9}\rm{Hz} $      \\
Information contained in every triple     & $4R$          & $4\times32\rm{bit}$             \\ 
The number of antenna     & $L$          & $4$             \\
The number of users     & $K$          & $5$             \\ \hline
\end{tabular}
\label{systemParameter}
\end{table}

\begin{figure*}[t]
  \centering
  \subfigure[System energy consumption versus Bandwidth]{
  \begin{minipage}[t]{0.48\linewidth}
    \centering
    \includegraphics[width=1\linewidth]{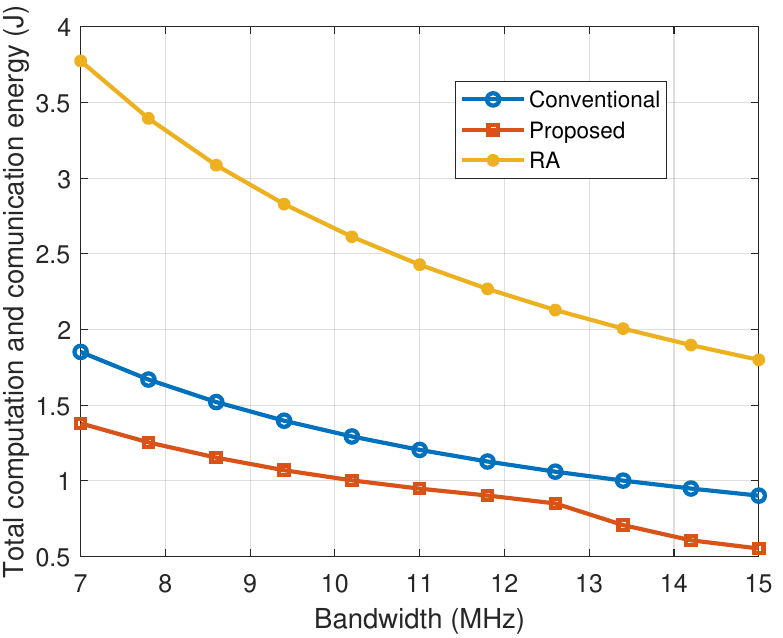}
    \label{pBandwidth}
    \vspace{0.02cm}
   \end{minipage}
   }
   \subfigure[System energy consumption versus Data size]{
  \begin{minipage}[t]{0.48\linewidth}
    \centering
    \includegraphics[width=1\linewidth]{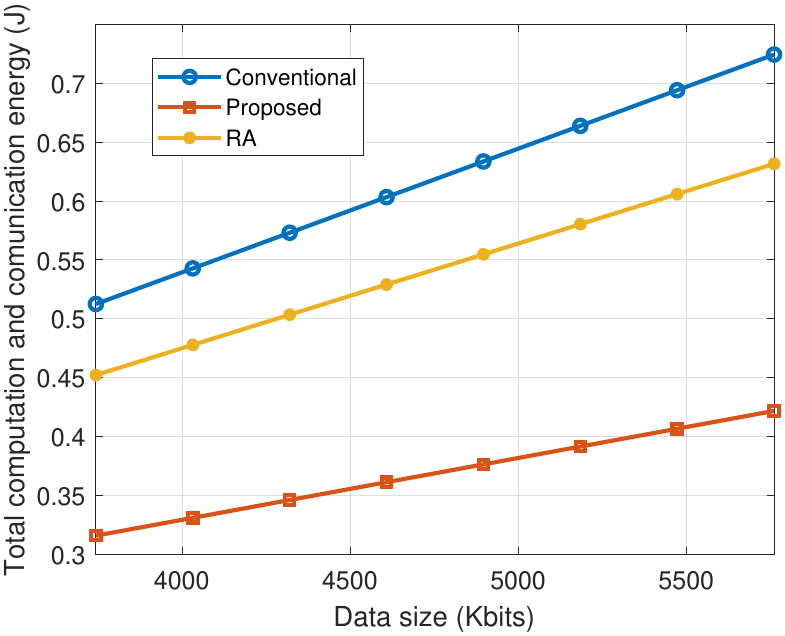}
    \label{pDataSize}
    \vspace{0.02cm}
   \end{minipage}
   }
\centering
\caption{Simulation results of power allocation and beamforming design sub-problem solving verification} 
\label{PowerAllocationSubProblem}
\end{figure*}

To verify the correctness of the outcomes of the sub-problems solving, we compare the obtained optimal solution to the sub-problems with the non-optimal schemes. The optimal solution to the sub-problem is labeled as `Proposed'. In Fig.~\ref{PowerAllocationSubProblem}, the baselines labeled as `Conventional' and `RA' mean based on the given maximum transmit power limit, the maximum power is evenly and randomly distributed to the transmission of shared information and the transmission of $K$ users' private information, respectively. Fig.~\ref{pBandwidth} shows that the changes in system energy consumption corresponding to the three allocation methods as the bandwidth changes, and Fig.~\ref{pDataSize} sets the variable as the total amount of data to be transmitted. It can be found that no matter which variable is, the allocation method that has been optimized always corresponds to the minimum system energy consumption under the same conditions.

\begin{figure*}[t]
  \centering
  \subfigure[System energy consumption versus Bandwidth]{
  \begin{minipage}[t]{0.48\linewidth}
    \centering
    \includegraphics[width=1\linewidth]{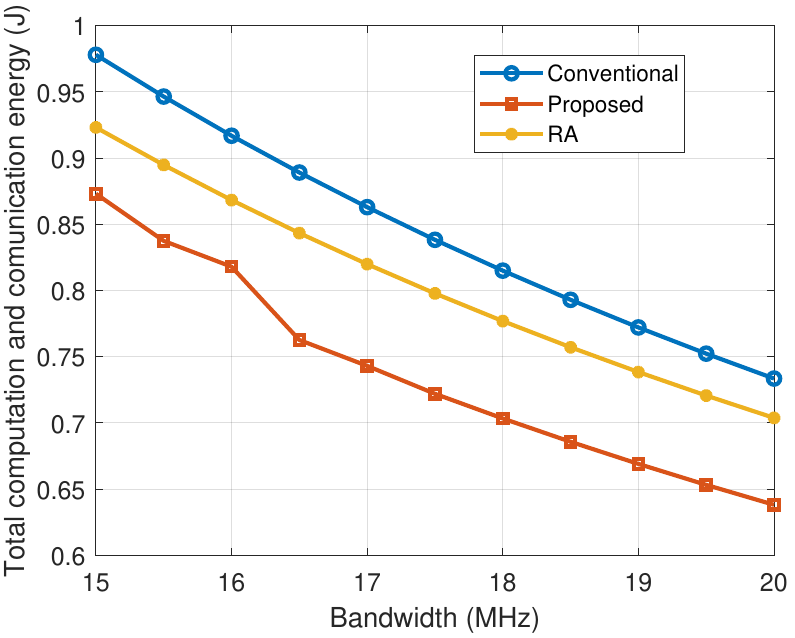}
    \label{OmegaBandwidth}
    \vspace{0.02cm}
   \end{minipage}
   }
   \subfigure[System energy consumption versus Data size]{
  \begin{minipage}[t]{0.48\linewidth}
    \centering
    \includegraphics[width=1\linewidth]{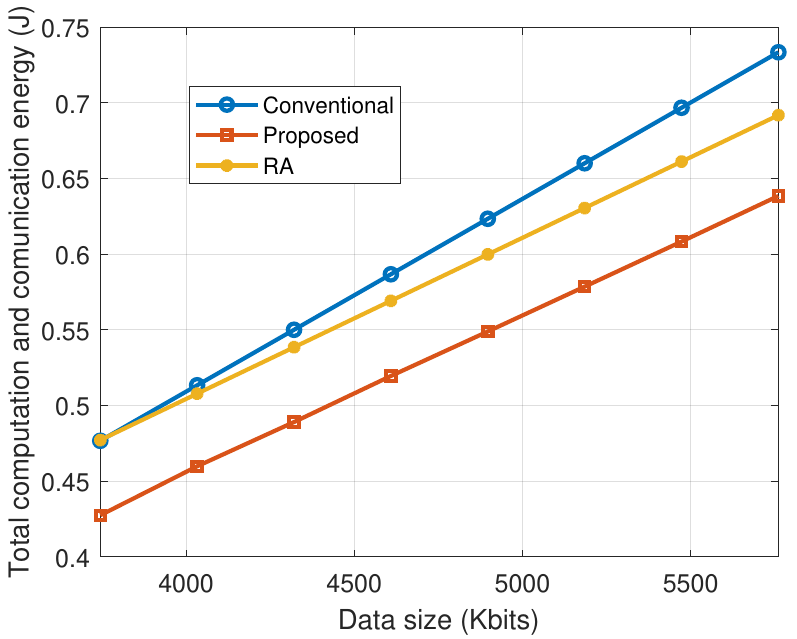}
    \label{OmegaDataSize}
    \vspace{0.02cm}
   \end{minipage}
   }
\centering
\caption{Simulation results of compression ratio sub-problem solving verification} 
\label{CompressionRatioSubProblem}
\end{figure*}

In terms of the compression ratio sub-problem, Fig.~\ref{CompressionRatioSubProblem} demonstrates the simulation results. The proposed optimization method is labeled as `Proposed', the baseline labeled as `Conventional' means during the transmission task, PSC model is not introduced, and the baseline `RA' means under the premise of satisfying the constraints of the sub-problem, the transmission of shared messages and the private messages of $K$ users are randomly assigned semantic compression ratios. As Fig.~\ref{OmegaBandwidth} reveals, after solving the convex problem, the system energy consumption corresponding to `Proposed’ under the same conditions is always the lowest. The same story happens as Fig.~\ref{OmegaDataSize} shows. However, we can find that when the data size is rather small, the quantity relation about system consumption between `Conventional' and `RA' is not the same of when the data size is larger, i.e. `RA' consumes more energy than `Conventional'. It may happen owing to the reason that the power allocation, beamforming design and computation capacity allocation at this time are not optimal under the simulation conditions at this time, and there is an unreasonable random compression rate for each user and public information, which may make its performance worse than that without compression.

\begin{figure*}[t]
  \centering
  \subfigure[System energy consumption versus Bandwidth]{
  \begin{minipage}[t]{0.48\linewidth}
    \centering
    \includegraphics[width=1\linewidth]{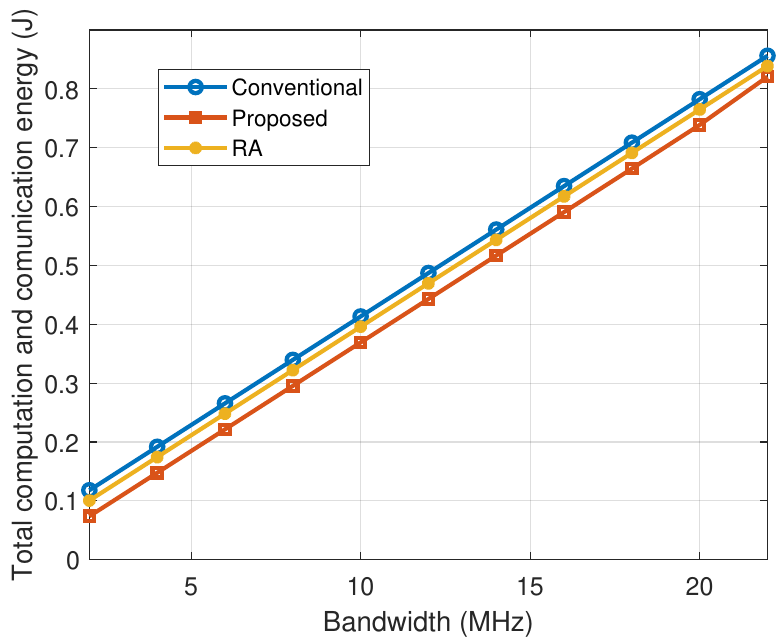}
    \label{fBandwidth}
    \vspace{0.02cm}
   \end{minipage}
   }
   \subfigure[System energy consumption versus Data size]{
  \begin{minipage}[t]{0.48\linewidth}
    \centering
    \includegraphics[width=1\linewidth]{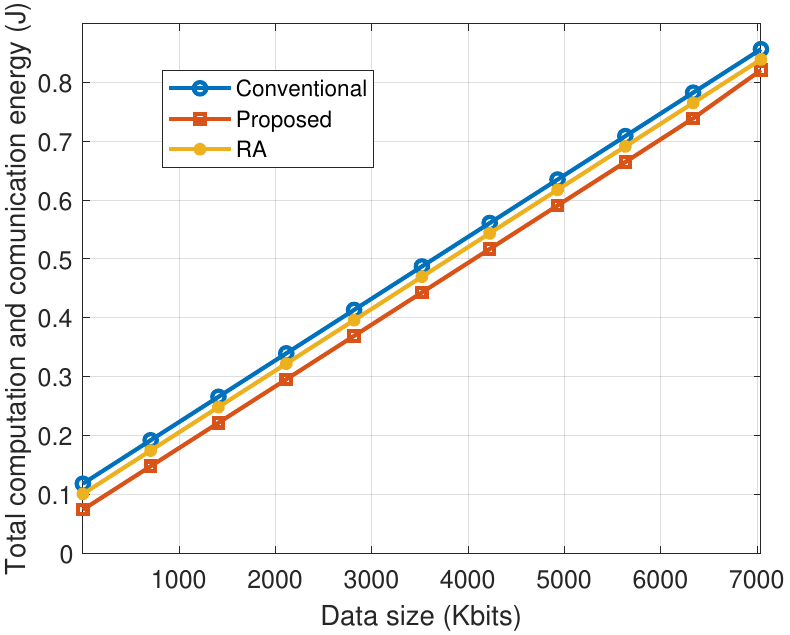}
    \label{fDataSize}
    \vspace{0.02cm}
   \end{minipage}
   }
\centering
\caption{Simulation results of computation capacity allocation sub-problem solving verification} 
\label{computationCapacitySubProblem}
\end{figure*}

Fig.~\ref{computationCapacitySubProblem} shows the outcomes of simulations on the topic of computation capacity allocation sub-problem. The baselines are set the same as in the Fig.~\ref{PowerAllocationSubProblem}. According to the derived result in \eqref{Computation_optimal3}, we can obtain the line `Proposed'. As the total amount of data increases, all three lines show an increasing trend, and the total energy consumption of the system corresponding to the optimal solution is less than that corresponding to the random allocation and average allocation ones.

\subsection{Optimization Problem Simulation Results and Analysis}



The proposed method is labeled as `RSMA'. In order to verify the superiority of the proposed probabilistic semantic compression model, we set up a control group, which does not use the compression model to send semantic information but directly transmits it, labeled as `Conventional'\cite{zhao2023semantic}. On the other hand, in order to prove the adaptability of RSMA to the proposed model, we set up different multiple access methods using SDMA and NOMA for comparison.

\begin{figure}[t]
\centering
\includegraphics[width=1\linewidth]{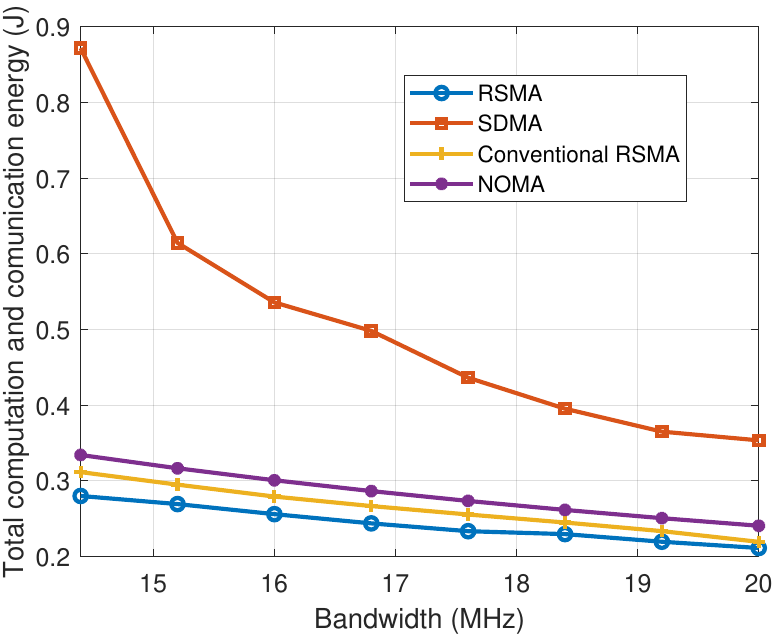}
\caption{Total computation and communication energy consumption vs. bandwidth of the system} 
\label{originalBandwidth}
\end{figure}

Fig.~\ref{originalBandwidth} illustrates that whether the data to be transmitted is compressed or not, the total energy consumption of the system using RSMA is always less than that of the system using SDMA and NOMA. In particular, when the available bandwidth of the total system is small, the energy consumption required by the SDMA method increases sharply, while the energy consumption required by the RSMA method still increases in a relatively smooth manner, proving that it is more appropriate to use RSMA to build a semantic communication transmission system. At the same time, by comparing the two lines of `Conventional' and `RSMA', it can be acknowledged that after PKG model is performed, the semantic communication system that jointly optimizes computation and communication can achieve lower energy consumption under the same bandwidth conditions.

\begin{figure}[t]
\centering
\includegraphics[width=1\linewidth]{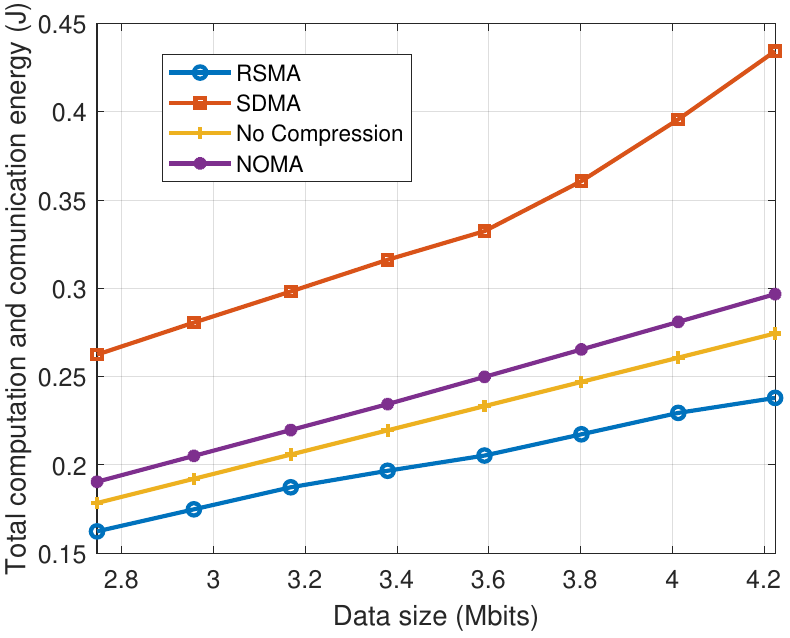}
\caption{Total computation and communication energy consumption vs. data size} 
\label{originaldataSize}
\end{figure}

Fig.~\ref{originaldataSize} presents the results on the total computation and communication energy versus data size varies. According to the figure, as the amount of data to be transmitted increases, the total energy consumption of the system corresponding to the four curves continues to increase. Among them, the energy consumption corresponding to the SDMA method is always the highest, followed by the NOMA method. The performance of the curve that uses the RSMA method for communication but does not use the semantic compression model proposed in the paper is rather favorable, and the proposed model has the best performance in the transmission task.
 
\section{Conclusion} \label{Conclusion}
In this paper, we have developed the PSC system from the single user scenario to multiuser scenario with RSMA. After sequentially modeling the representation, compression, transmission and recovering of multiuser semantic information, we constructed an optimization problem to minimize the energy consumption of the system on the basis of satisfying certain optimization constraints. An alternating optimization algorithm is then proposed to solve the energy-minimize problem. Simulation results indicate that the proposed system exhibits a better performance than the model using no semantic compression and the models adapting other multiple access methods.

\bibliographystyle{IEEEtran}
\bibliography{MMM}

\begin{thebibliography}{10}
\providecommand{\url}[1]{#1}
\csname url@samestyle\endcsname
\providecommand{\newblock}{\relax}
\providecommand{\bibinfo}[2]{#2}
\providecommand{\BIBentrySTDinterwordspacing}{\spaceskip=0pt\relax}
\providecommand{\BIBentryALTinterwordstretchfactor}{4}
\providecommand{\BIBentryALTinterwordspacing}{\spaceskip=\fontdimen2\font plus
\BIBentryALTinterwordstretchfactor\fontdimen3\font minus \fontdimen4\font\relax}
\providecommand{\BIBforeignlanguage}[2]{{%
\expandafter\ifx\csname l@#1\endcsname\relax
\typeout{** WARNING: IEEEtran.bst: No hyphenation pattern has been}%
\typeout{** loaded for the language `#1'. Using the pattern for}%
\typeout{** the default language instead.}%
\else
\language=\csname l@#1\endcsname
\fi
#2}}
\providecommand{\BIBdecl}{\relax}
\BIBdecl

\bibitem{zhang2023model}
P.~Zhang, X.~Xu, C.~Dong, K.~Niu, H.~Liang, Z.~Liang, X.~Qin, M.~Sun, H.~Chen, N.~Ma \emph{et~al.}, ``Model division multiple access for semantic communications,'' \emph{Frontiers of Information Technology \& Electronic Engineering}, vol.~24, no.~6, pp. 801--812, 2023.

\bibitem{shannon1948mathematical}
C.~E. Shannon, ``A mathematical theory of communication,'' \emph{The Bell system technical journal}, vol.~27, no.~3, pp. 379--423, 1948.

\bibitem{shannon1949mathematical}
C.~E. Shannon and W.~Weaver, \emph{The mathematical theory of communication, by CE Shannon (and recent contributions to the mathematical theory of communication), W. Weaver}.\hskip 1em plus 0.5em minus 0.4em\relax University of illinois Press, 1949.

\bibitem{weaver1953recent}
W.~Weaver, ``Recent contributions to the mathematical theory of communication,'' \emph{ETC: a review of general semantics}, pp. 261--281, 1953.

\bibitem{xu2023edge}
W.~Xu, Z.~Yang, D.~W.~K. Ng, M.~Levorato, Y.~C. Eldar, and M.~Debbah, ``Edge learning for b5g networks with distributed signal processing: Semantic communication, edge computing, and wireless sensing,'' \emph{IEEE journal of selected topics in signal processing}, vol.~17, no.~1, pp. 9--39, 2023.

\bibitem{uysal2022semantic}
E.~Uysal, O.~Kaya, A.~Ephremides, J.~Gross, M.~Codreanu, P.~Popovski, M.~Assaad, G.~Liva, A.~Munari, B.~Soret \emph{et~al.}, ``Semantic communications in networked systems: A data significance perspective,'' \emph{IEEE Network}, vol.~36, no.~4, pp. 233--240, 2022.

\bibitem{kountouris2021semantics}
M.~Kountouris and N.~Pappas, ``Semantics-empowered communication for networked intelligent systems,'' \emph{IEEE Communications Magazine}, vol.~59, no.~6, pp. 96--102, 2021.

\bibitem{xie2021deep}
H.~Xie, Z.~Qin, G.~Y. Li, and B.-H. Juang, ``Deep learning enabled semantic communication systems,'' \emph{IEEE Transactions on Signal Processing}, vol.~69, pp. 2663--2675, 2021.

\bibitem{xie2020lite}
H.~Xie and Z.~Qin, ``A lite distributed semantic communication system for internet of things,'' \emph{IEEE Journal on Selected Areas in Communications}, vol.~39, no.~1, pp. 142--153, 2020.

\bibitem{zhang2010understanding}
Y.~Zhang, R.~Jin, and Z.-H. Zhou, ``Understanding bag-of-words model: a statistical framework,'' \emph{International journal of machine learning and cybernetics}, vol.~1, pp. 43--52, 2010.

\bibitem{wang2019evaluating}
B.~Wang, A.~Wang, F.~Chen, Y.~Wang, and C.-C.~J. Kuo, ``Evaluating word embedding models: Methods and experimental results,'' \emph{APSIPA transactions on signal and information processing}, vol.~8, p. e19, 2019.

\bibitem{vayansky2020review}
I.~Vayansky and S.~A. Kumar, ``A review of topic modeling methods,'' \emph{Information Systems}, vol.~94, p. 101582, 2020.

\bibitem{bordes2014semantic}
A.~Bordes, X.~Glorot, J.~Weston, and Y.~Bengio, ``A semantic matching energy function for learning with multi-relational data: Application to word-sense disambiguation,'' \emph{Machine learning}, vol.~94, pp. 233--259, 2014.

\bibitem{wang2017knowledge}
Q.~Wang, Z.~Mao, B.~Wang, and L.~Guo, ``Knowledge graph embedding: A survey of approaches and applications,'' \emph{IEEE transactions on knowledge and data engineering}, vol.~29, no.~12, pp. 2724--2743, 2017.

\bibitem{jiang2022reliable}
S.~Jiang, Y.~Liu, Y.~Zhang, P.~Luo, K.~Cao, J.~Xiong, H.~Zhao, and J.~Wei, ``Reliable semantic communication system enabled by knowledge graph,'' \emph{Entropy}, vol.~24, no.~6, p. 846, 2022.

\bibitem{wang2023knowledge}
B.~Wang, R.~Li, J.~Zhu, Z.~Zhao, and H.~Zhang, ``Knowledge enhanced semantic communication receiver,'' \emph{IEEE Communications Letters}, 2023.

\bibitem{dai2022nonlinear}
J.~Dai, S.~Wang, K.~Tan, Z.~Si, X.~Qin, K.~Niu, and P.~Zhang, ``Nonlinear transform source-channel coding for semantic communications,'' \emph{IEEE Journal on Selected Areas in Communications}, vol.~40, no.~8, pp. 2300--2316, 2022.

\bibitem{dong2022semantic}
C.~Dong, H.~Liang, X.~Xu, S.~Han, B.~Wang, and P.~Zhang, ``Semantic communication system based on semantic slice models propagation,'' \emph{IEEE Journal on Selected Areas in Communications}, vol.~41, no.~1, pp. 202--213, 2022.

\bibitem{wang2022wireless}
S.~Wang, J.~Dai, Z.~Liang, K.~Niu, Z.~Si, C.~Dong, X.~Qin, and P.~Zhang, ``Wireless deep video semantic transmission,'' \emph{IEEE Journal on Selected Areas in Communications}, vol.~41, no.~1, pp. 214--229, 2022.

\bibitem{jiang2022wireless}
P.~Jiang, C.-K. Wen, S.~Jin, and G.~Y. Li, ``Wireless semantic communications for video conferencing,'' \emph{IEEE Journal on Selected Areas in Communications}, vol.~41, no.~1, pp. 230--244, 2022.

\bibitem{bao2011towards}
J.~Bao, P.~Basu, M.~Dean, C.~Partridge, A.~Swami, W.~Leland, and J.~A. Hendler, ``Towards a theory of semantic communication,'' in \emph{2011 IEEE Network Science Workshop}.\hskip 1em plus 0.5em minus 0.4em\relax IEEE, 2011, pp. 110--117.

\bibitem{maatouk2022age}
A.~Maatouk, M.~Assaad, and A.~Ephremides, ``The age of incorrect information: An enabler of semantics-empowered communication,'' \emph{IEEE Transactions on Wireless Communications}, vol.~22, no.~4, pp. 2621--2635, 2022.

\bibitem{weng2021semantic}
Z.~Weng and Z.~Qin, ``Semantic communication systems for speech transmission,'' \emph{IEEE Journal on Selected Areas in Communications}, vol.~39, no.~8, pp. 2434--2444, 2021.

\bibitem{guler2018semantic}
B.~G{\"u}ler, A.~Yener, and A.~Swami, ``The semantic communication game,'' \emph{IEEE Transactions on Cognitive Communications and Networking}, vol.~4, no.~4, pp. 787--802, 2018.

\bibitem{xu2024resource}
R.~Xu, Z.~Yang, Z.~Zhao, Q.~Yang, and Z.~Zhang, ``Resource allocation for green probabilistic semantic communication with rate splitting,'' \emph{arXiv preprint arXiv:2404.00612}, 2024.

\bibitem{mao2018rate}
Y.~Mao, B.~Clerckx, and V.~O. Li, ``Rate-splitting multiple access for downlink communication systems: Bridging, generalizing, and outperforming sdma and noma,'' \emph{EURASIP journal on wireless communications and networking}, vol. 2018, pp. 1--54, 2018.

\bibitem{9831440}
Y.~Mao, O.~Dizdar, B.~Clerckx, R.~Schober, P.~Popovski, and H.~V. Poor, ``Rate-splitting multiple access: Fundamentals, survey, and future research trends,'' \emph{IEEE Communications Surveys \& Tutorials}, vol.~24, no.~4, pp. 2073--2126, 2022.

\bibitem{wang2022performance}
Y.~Wang, M.~Chen, T.~Luo, W.~Saad, D.~Niyato, H.~V. Poor, and S.~Cui, ``Performance optimization for semantic communications: An attention-based reinforcement learning approach,'' \emph{IEEE Journal on Selected Areas in Communications}, vol.~40, no.~9, pp. 2598--2613, 2022.

\bibitem{lobo1998applications}
M.~S. Lobo, L.~Vandenberghe, S.~Boyd, and H.~Lebret, ``Applications of second-order cone programming,'' \emph{Linear algebra and its applications}, vol. 284, no. 1-3, pp. 193--228, 1998.

\bibitem{zhao2023semantic}
Z.~Zhao, Z.~Yang, Q.-V. Pham, Q.~Yang, and Z.~Zhang, ``Semantic communication with probability graph: A joint communication and computation design,'' in \emph{2023 IEEE 98th Vehicular Technology Conference (VTC2023-Fall)}.\hskip 1em plus 0.5em minus 0.4em\relax IEEE, 2023, pp. 1--5.

\end{thebibliography}

\end{document}